\documentclass{PoS}
\usepackage{amsfonts}
\usepackage{amssymb}
\usepackage[mathscr]{eucal}
\usepackage{amsfonts,amssymb,amsthm}

\def \be{\begin{equation}}
\def \ee{\end{equation}}
\def \bes{\begin{eqnarray}}
\def \ees{\end{eqnarray}}

\newcommand{\Cb}{{\rm \bf C}}

\def \sl2{SL(2,\Cb)}
\title{Group field theory as the microscopic description of the quantum spacetime fluid: a new perspective on the continuum in quantum gravity}
\ShortTitle{Group field theory as the microscopic description of
the quantum spacetime fluid}
\author{\speaker{Daniele Oriti}%
        \\
       Institute for Theoretical Physics, Utrecht University\\
       E-mail: \email{d.oriti@phys.uu.nl}}
\abstract{We introduce the group field theory (GFT) formalism for
non-perturbative quantum gravity, and present it as a potential
unifying framework for several other quantum gravity approaches,
i.e. loop quantum gravity and simplicial quantum gravity ones. We
then argue in favor of and present in detail what we believe is a
new GFT perspective on the emergence of continuum spacetime from
discrete quantum structures, based on the idea of quantum space as
a condensed matter system. We put forward a more specific, albeit
still very much tentative, proposal for the relevant phase of the
GFT corresponding to the continuum: a Bose-Einstein condensate of
GFT quanta. Finally, we sketch how the proposal may be realised
and its effective dynamics could be extracted in the GFT setting
and compared with continuum gravity theories.}

\FullConference{From Quantum to Emergent Gravity: Theory and Phenomenology\\
                June 11-15 2007\\
                Trieste, Italy}

\begin{document}

\section{Introduction}
The purpose of this contribution to the debate on \lq\lq Quantum
and emergent gravity\rq\rq is fourfold.

First of all, we would like to introduce the group field theory
(GFT) formalism \cite{iogft, iogft2,laurentgft}, that has recently
attracted interest in the general area of non-perturbative quantum
gravity, and is currently mainly used in the context of Loop
Quantum Gravity \cite{LQG}. We will describe the general features
of the formalism, at both kinematical and dynamical level, and
provide an interpretation for them.

Second, we would like to portrait a picture of group field
theories as a common framework and a unifying language for several
approaches to quantum gravity, in particular loop quantum gravity
and simplicial quantum gravity (i.e. quantum Regge calculus and
dynamical triangulations), by sketching how the basic ingredients
of these various approaches can be identified within the GFT
setting. We will argue that the pictures of quantum spacetime,
developed in the various approaches, are compatible and can help
completing each other, while acquiring a new interpretation within
the GFT framework. GFTs can then represent a suitable context in
which all these different approaches can inform, cross-fertilize
and improve each other with the achieved results and insights into
the nature of quantum geometry, and with the tools they have
developed to study it. In doing so, of course, we will discuss why
we think is useful to move from the contexts provided by each of
these quantum gravity approaches to the GFT one.

Third, we want to stress the need to devote our research efforts
to tackle the issue of the continuum approximation of the quantum
discrete structures that these various approaches identify as the
fundamental building blocks of spacetime. Only if we are able to
show convincingly that a good continuum description of spacetime,
with its dynamics governed by (some modified version of) General
Relativity, emerges naturally from the formulation of quantum
gravity we favor, we will have a truly convincing argument for
believing this formulation. This is of course well-known by
researchers working in non-perturbative quantum gravity, and in
particular in the approaches we have just mentioned: loop quantum
gravity (and spin foam models), quantum Regge calculus and
(causal) dynamical triangulations. Indeed, many techniques and
strategies have been developed, within these various approaches,
to solve the continuum (and semi-classical) riddle, and many
results already obtained. We will briefly discuss, and try to
re-phrase, them in the GFT language. This will allow us to both
understand them as providing insights about different regimes and
features of the same type of models, and clarify in which sense
they do not represent the most convenient or natural way to
approach the continuum problem from a GFT point of view.

Last, we will argue that group field theories offer new and
powerful tools to tackle the problem of the continuum in quantum
gravity, together with a new perspective on the whole issue, that
could prove decisive for settling it, at the same time developing
further and going beyond the insights obtained from the other
approaches mentioned above. The suggestion will basically be that
we could try to view spacetime as a (peculiar indeed) condensed
matter system, with the GFT representing the microscopic
description of its \lq\lq atoms\rq\rq, and providing the starting
point for studying both the statistical mechanics and the
effective dynamics of large number of them, which we will
tentatively identify with continuum physics. In particular, group
field theories can offer the context and the tools to realize
explicitly the intriguing idea of spacetime as a condensate of
fundamental building blocks and of continuum geometry as an
emergent concept. We will then put forward a proposal for this GFT
condensate, suggest some concrete research directions (some of
which currently pursued), and offer some speculation on how a
continuum spacetime and General Relativity can emerge in this
scheme, again making use (also) of the condensed matter analogy.

Given its aims, this article will contain a limited amount of
technicalities, only those needed to introduce the main GFT idea
and general formalism, and only references to and brief
discussions of the many results obtained both in the GFT context
and in the context of the other approaches to quantum gravity we
will mention. At the same time, it may contain a more than average
amount of speculations, especially in its last part, when we will
try to forecast where the new perspective we are advocating may
lead to. We will hopefully compensate for this by trying to be as
precise as possible in presenting the main ideas, motivations and
arguments behind this perspective, and to convince the reader that
this may be an intriguing and reasonable picture of what recent
results in quantum gravity research are pointing to.

\section{The group field theory formalism}
We now proceed to introduce the main features of the GFT
formalism. We refer to the literature, in particular the reviews
\cite{iogft,iogft2,laurentgft}, for a more complete and detailed
treatment and a more extensive list of references.
\subsection{Kinematics: the fundamental building blocks of quantum space}
We start from a field taken to be a
  $\mathbb{C}$-valued function of D group elements, for a generic
  group $G$, one for each of the D boundary (D-2)-faces of the
  (D-1)-simplex that the field $\phi$ represents: $$\phi(g_1,g_2,...,g_D):
  G^{\times D}\rightarrow \mathbb{C}.$$

In models (aiming at) describing D-dimensional quantum gravity,
this field is interpreted as a second quantized (D-1)-simplex,
with (D-2)-faces of the same labelled by group theoretic data,
interpreted as (pre-)geometric elementary quantities, or discrete
quantum gravity variables. Equivalently, the same data can be
associated to the links of a topologically dual graph, and the
field is then seen as the second quantization of a spin network
functional \cite{LQG}. This means that GFTs can be seen
equivalently as a second quantized formulation of spin network
dynamics or as a field theory {\it of} simplicial geometry. We can
identify the ordering of the arguments of the field with a choice
of orientation for the (D-1)-simplex it represents, and we require
invariance of the field under even permutations $\sigma$ of its
arguments and trade odd permutations with complex conjugation of
the field. Other symmetry properties can also be considered. An
additional symmetry that is usually imposed on the field is the
invariance under diagonal action of the group $G$ on the D
arguments of the field: $\phi(g_1,...,g_D)=\phi(g_1g,...,g_Dg)$;
but this is again model-dependent, of course, and in the models of
\cite{generalised,iotimgft}, for example, only invariance under a
certain proper subgroup is imposed. This is the simplicial
counterpart of the Lorentz gauge invariance of continuum and
discrete first order gravity actions, and it has also the
geometric interpretation, at the simplicial level, of requiring
the D faces of a (D-1)-simplex to close.

A momentum representation for the field and its dynamics is
obtained by harmonic analysis on the group manifold $G$. The field
can be expanded in modes as:
$$\phi(g_i)=\sum_{J_i,\Lambda,k_i}\phi^{J_i\Lambda}_{k_i}\left( \prod_iD^{J_i}_{k_il_i}(g_i)\right) C^{J_1..J_D\Lambda}_{l_1..l_D}, $$ with the $J$'s
labelling representations of $G$, the $k$'s vector indices in the
representation spaces, and the $C$'s being intertwiners of the
group $G$. We have labelled an orthonormal basis of intertwiners
by an extra parameter $\Lambda$ (depending on the group chosen and
on the dimension D, this may actually be a shorthand notation for
{\it a set} of parameters). That this decomposition is possible is
not guaranteed in general, but it is in fact true for all the
known quantum gravity GFT models, which are based on the Lorentz
group or on extensions of it. The proper geometric interpretation
of the field variables can be identified by looking at the Feynman
amplitudes for the GFT at hand, that either have the form of
discrete path integrals for some gravity action
\cite{generalised,iotimgft} or can be derived from one
\cite{iogft,iogft2,laurentgft}. This interpretation depends of
course on the specific model considered. However, generally
speaking, the group variables are seen to represent parallel
transport of a (gravity) connection along elementary paths dual to
the (D-2)-faces, and the representations $J$ are usually put in
correspondence with the volumes of the same (D-2)-faces.

\begin{figure}[t]
\includegraphics[width=15.4cm, height=3cm]{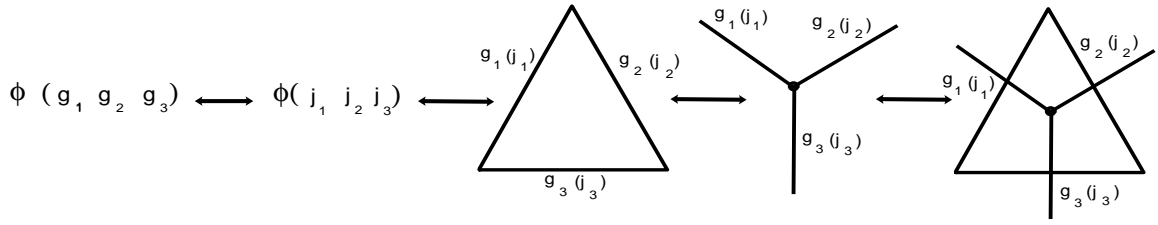}
\caption{For the $D=3$ case, the association of a field with a
2-simplex, or equivalently its dual vertex, and of its arguments
with the 1-faces of it, or equivalently with the links incident to
the vertex, together with the labelling by group-theoretic
variables.}
\end{figure}

Just as one identifies a single field with a single (D-1)-simplex,
a simplicial space built out of $N$ such (D-1)-simplices is
described by a suitable polynomial in the field variables, with
constraints among the group or representation data, implementing
the fact that some of their (D-2)-faces are identified. For
example, a state describing two (D-1)-simplices glued along one
common (D-2)-face would be represented by: $\phi^{J_1
J_2..J_D\Lambda}_{k_1 k_2...k_D} \phi^{\tilde{J}_1
  J_2...\tilde{J}_D\tilde{\Lambda}}_{\tilde{k}_1 k_2...\tilde{k}_D}$,
where the gluing is along the face labelled by the representation
$J_2$, and effected by the contraction of the corresponding vector
indices (of course, states corresponding to disjoint
(D-1)-simplices are also allowed).

\begin{figure}
\includegraphics[width=12.8cm, height=3cm]{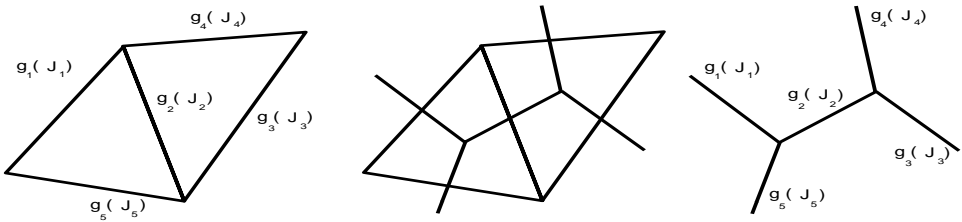}

\caption{A \lq 2-particle state\rq (again, in the D=3 example)}
\end{figure}

We see that states of the theory are then labelled, in momentum
space, by {\it spin networks} based on the group $G$ \cite{LQG}.

GFT observables are given \cite{laurentgft} by gauge invariant
functionals of the GFT field, and can be constructed in momentum
space using again spin networks according to the formula:
$$ O_{\Psi=(\gamma, j_e,i_v)}(\phi)=\left(\prod_{(ij)}\int dg_{ij}dg_{ji}\right) \Psi_{(\gamma, j_e,i_v)}(g_{ij}g_{ji}^{-1})\prod_i \phi(g_{ij}),$$
where  $\Psi_{(\gamma, j_e,i_v)}(g)$ identifies a spin network
functional \cite{LQG} for the spin network labelled by a graph
$\gamma$ with representations $j_e$ associated to its edges and
intertwiners $i_v$ associated to its vertices, and $g_{ij}$ are
group elements associated to the edges $(ij)$ of $\gamma$ that
meet at the vertex $i$.

Thus, {\bf group field theories describe a quantum space in terms
of fundamental building blocks, the quanta of the GFT field, that
acquire then the status of \lq\lq atoms of space\rq\rq in this
setting, and that can be represented both as spin network vertices
or as elementary (D-1)-simplices. A generic quantum state will be
a \lq\lq many-particle\rq\rq configuration for these quanta,
representing some extended discrete structure (a larger spin
network or a larger (D-1)-triangulation)} characterized by both
the \lq\lq particle number\rq\rq and by additional symmetries or
constraints imposed, specifying how the fundamental building
blocks are glued together. This picture can be made more precise
and a Fock space characterization of the GFT state space (and thus
of quantum space, in this framework) can be obtained after
Hamiltonian analysis of specific GFT models \cite{iojimmy}.

\subsection{Dynamics: the interaction and evolution of the atoms of space}
On the basis of the above kinematical structure, one aims at
defining a field theory for describing the interaction of
fundamental atoms of space, and in which {\bf a typical
interaction process will be characterized by a D-dimensional
simplicial complex. In the dual picture, the same will be
represented as a spin foam (labelled 2-complex).} This is the
straightforward generalization of the way in which 2d discretized
surfaces emerge from the interaction of matrices (graphically,
segments)\cite{mm}, or ordinary Feynman graphs emerge from the
interaction of point particles. A {\it discrete} spacetime emerge
then from the theory as a virtual construct, a possible
interaction process among the GFT quanta.

In order for this to be realized, the classical field action in
group field theories has to be chosen appropriately. In this
choice lies the main peculiarity of GFTs with respect to ordinary
field theories. This action, in configuration space, has the
general structure:

\begin{eqnarray} \hspace{-0.4cm} S_D(\phi, \lambda)= \frac{1}{2}\left(\prod_{i=1}^D\int
  dg_id\tilde{g}_i\right)
  \phi(g_i)\mathcal{K}(g_i\tilde{g}_i^{-1})\phi(\tilde{g}_i)
  +
  \frac{\lambda}{(D+1)!}\left(\prod_{i\neq j =1}^{D+1}\int dg_{ij}\right)
  \phi(g_{1j})...\phi(g_{D+1 j})\,\mathcal{V}(g_{ij}g_{ji}^{-1})\;\;, \label{eq:action}
\end{eqnarray}
and it is of course the choice of kinetic and interaction
functions $\mathcal{K}$ and $\mathcal{V}$ that define the specific
model considered. Obviously, the same action can be written in
momentum space after harmonic decomposition on the group manifold.
The interaction term describes the interaction of D+1
(D-1)-simplices to form a D-simplex (\lq a fundamental virtual
spacetime event\rq) by gluing along their (D-2)-faces (arguments
of the fields), that are {\it pairwise} linked by the interaction
vertex. The nature of this interaction is specified by the choice
of function $\mathcal{V}$. The kinetic term involves two fields
each representing a given (D-1)-simplex seen from one of the two
D-simplices (interaction vertices) sharing it, so that the choice
of kinetic functions $\mathcal{K}$ specifies how the information
and therefore the geometric degrees of freedom corresponding to
their D (D-2)-faces are propagated from one vertex of interaction
to another. One can consider generalizations of the above
combinatorial structure, corresponding to the gluing of
(D-1)-simplices to form different sorts of D-dimensional complexes
(e.g. hypercubes etc).

Some examples of GFT actions are: 1) those corresponding to the
kinetic and vertex functions:

\be \mathcal{K}(g_i,\tilde{g}_i) = \prod_{i=1}^{D}
\delta(g_i\tilde{g}_i^{-1}),\;\;\;\;\;\;\mathcal{V}(g_{ij},g_{ji})
=  \prod_{i<j=1}^{D+1}\delta(g_{ij}g_{ji}^{-1}), \label{BF} \ee

which produce a perturbative quantum dynamics that can be related
to topological BF theories in any dimension, for internal gauge
group $G$; 2) models in which suitably defined additional
constraints on the same BF-type kinetic and/or vertex terms are
imposed, and which aim at representing the GFT equivalent of the
constraint reducing BF theory to gravity in a Plebanski-like
formulation of the same \cite{DP-F-K-R,P-R,iolaurentkirill}; 3)
extended models based on more than Lorentz group variables and
characterized by a proper differential operator playing the role
of kinetic term, one example of which is the class of models in
\cite{iotimgft}, using a complex field on $(G\times X)^D$, with
$G$ being the Lorentz group and $X$ a metric space isomorphic to
the Lie algebra of $G$, and based on the kinetic and vertex terms:

\begin{eqnarray}
\hspace{-0.5cm} \mathcal{K}(g_i,x_i, \tilde{g}_i,\tilde{x}_i) = \,
\prod_i\,\left( \triangle_{i} + \square_i \right)\delta(g_i
\tilde{g}_{i}^{-1})\delta(x_i - \tilde{x}_{i}^{-1}) \;\;\;\;\;\;
\mathcal{V}(g_{ij},x_{ij}) =\,\prod_{i\neq j}^{} \delta(g_{ij}
g_{ji}^{-1})\delta(x_{ij} - x_{ji}) \label{EF}
\end{eqnarray}
where $g_i\in G$, $x_i\in X$, $\triangle$ is the Laplace-Beltrami
on $X$ and $\square$ is the Laplace-Beltrami on $G$; these last
models produce Feynman amplitudes with the interpretation of
simplicial path integrals for 1st order gravity actions
\cite{iotimgft}.

\ \

Let us now turn to the quantum dynamics. Most of the research in
this area has concerned the perturbative aspects of this dynamics
around the no-particle state, the complete vacuum, and the main
guide for model building have been, up to now, only the properties
of the resulting Feynman amplitudes:

$$ Z\,=\,\int
\mathcal{D}\phi\,e^{iS[\phi]}\,=\,\sum_{\Gamma}\,\frac{\lambda^{N_v(\Gamma)}}{sym[\Gamma]}\,Z(\Gamma),
$$
where $N_v$ is the number of interaction vertices $v$ in the
Feynman diagram $\Gamma$, $sym[\Gamma]$ is the number of
automorphisms of $\Gamma$ and $Z(\Gamma)$ the corresponding
Feynman amplitude. Each edge of the Feynman graph is made of $D$
strands, one for each argument of the field and each one is then
re-routed at the interaction vertex, with the combinatorial
structure of an $D$-simplex, following the pairing of field
arguments in the vertex operator.

\begin{figure}[here]
\setlength{\unitlength}{1cm}
\begin{minipage}[t]{3.5cm}
\hspace{-0.3cm}\includegraphics[width=3.5cm,
height=2.5cm]{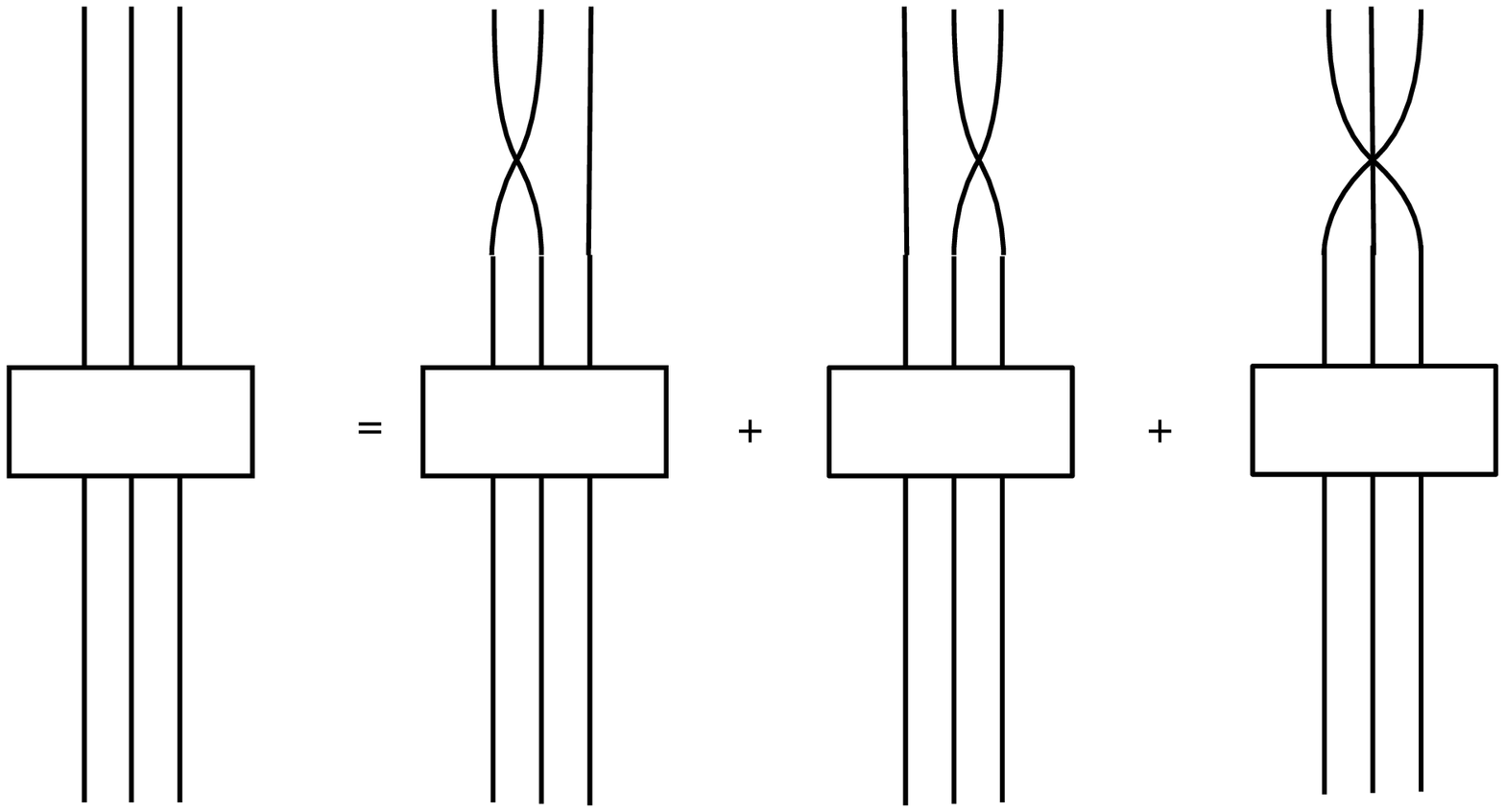}
\end{minipage}
\hspace{0.3cm}
\begin{minipage}[t]{5.5cm}
\includegraphics[width=3.5cm, height=2.5cm]{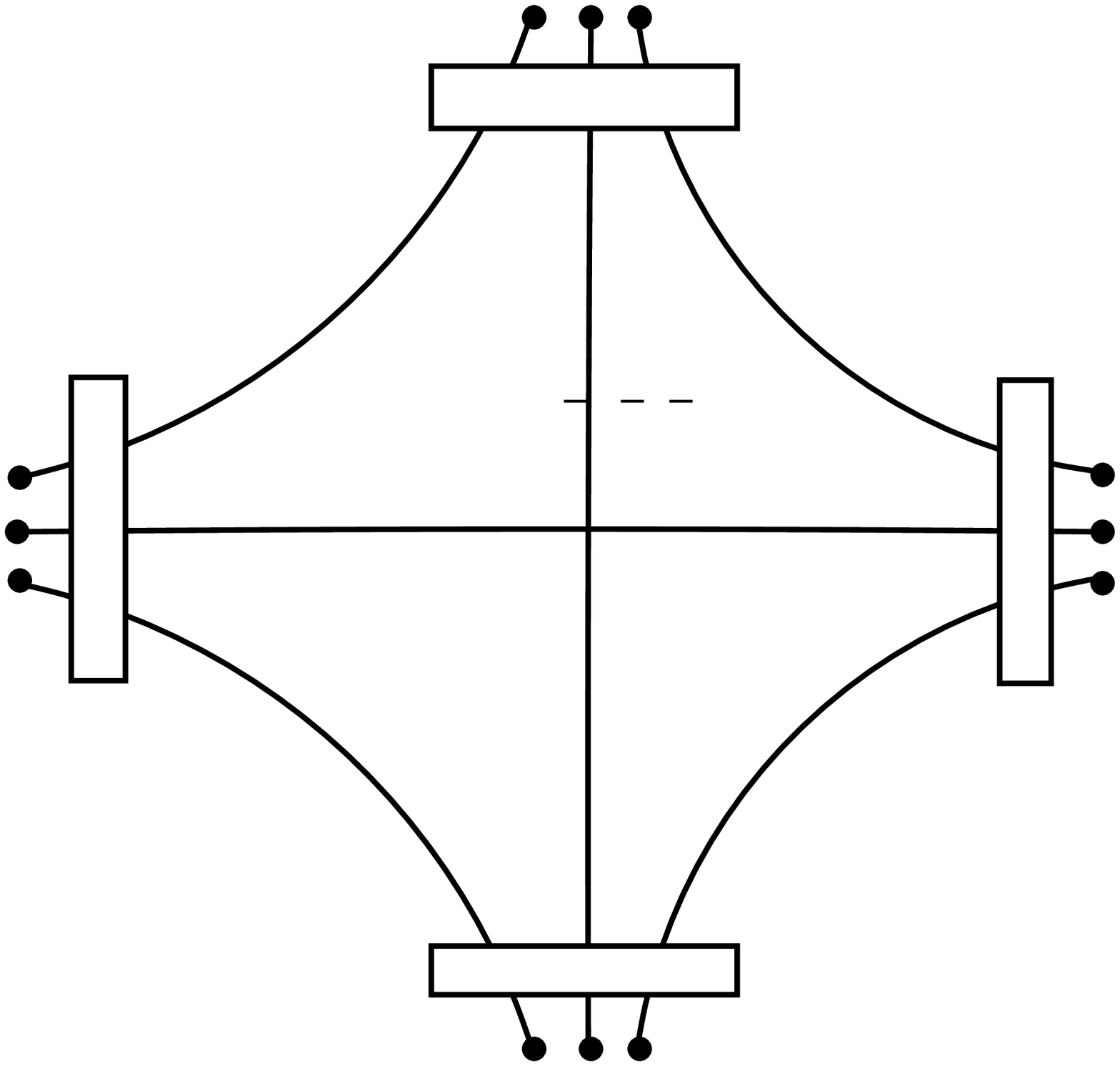}
\end{minipage}
\hspace{0.1cm}
\begin{minipage}[t]{5cm}
\includegraphics[width=7cm, height=3cm]{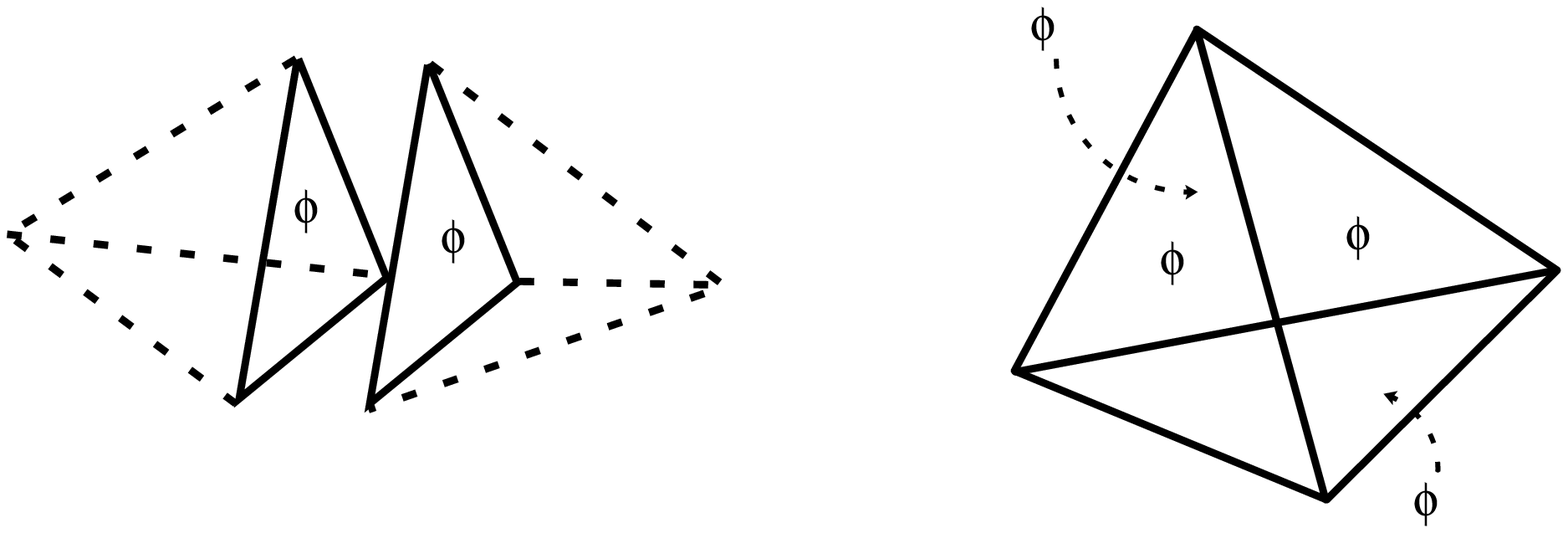}
\end{minipage}
\caption{The basic building blocks of the GFT Feynman diagrams
(for $D=3$).}
\end{figure}

Each strand in an edge of the Feynman diagram goes through several
vertices, coming back where it started, for closed Feynman
diagrams, and therefore identifies a 2-cell (for open graphs, it
may end up on the boundary, but still identifies a 2-cell). Each
Feynman diagram $\Gamma$ is then a collection of 2-cells, edges
and vertices, i.e. a 2-complex, that, because of the chosen
combinatorics for the arguments of the field in the action, is
topologically dual to a D-dimensional simplicial complex. Notice
that the resulting 2-cells can be glued (i.e. can share edges) in
all sorts of ways, forming for example \lq\lq bubbles\rq\rq, i.e.
closed 3-cells.

No restriction on the topology of the diagram/complex is imposed,
a priori, in the construction, so the resulting
complexes/triangulations can have arbitrary topology. Each of them
corresponds to a particular {\it scattering
  process} of the fundamental building blocks of space,
i.e. (D-1)-simplices/spin network vertices. Each line of
propagation, made as we said out of D strands, is labelled, on top
of the group/representation data, by a permutation of $(1,..,D)$,
representing the labelling of the field variables, and all these
data are summed over in the construction of the Feynman expansion.
The sum over permutations affects directly the combinatorics of
the allowed gluings of vertices with propagators\cite{DP-P}.
\begin{figure}[here]
\includegraphics[width=12.5cm, height=3cm]{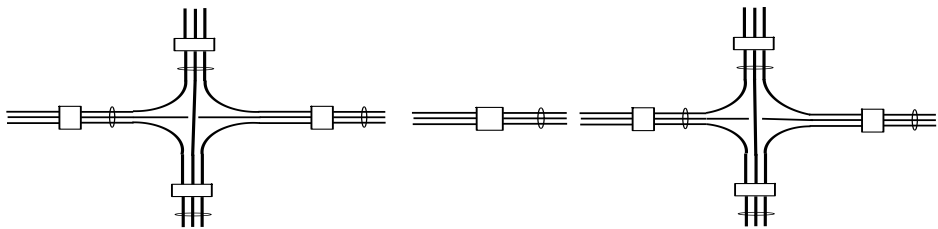}
\caption{The gluing of vertices of interaction through
propagators, again in the D=3 example. The rectangles represent
the additional integrations imposing gauge invariance under the
action of $G$, while the ellipses represent the implicit sum over
permutations of the (labels of the) strands to be glued.}
\end{figure}

As said, each strand in a propagation line carries a field
variable, e.g. a group element in configuration space or a
representation label in momentum space. After the closure of the
strand to form a 2-cell in a closed diagram, the same
representation label ends up being associated to this 2-cell.
Therefore in momentum space each Feynman graph is given by a spin
foam (a 2-complex with faces labelled by representation
variables), and each Feynman amplitude (a complex function of the
representation labels, obtained by contracting vertex amplitudes
with propagator functions) by a so-called spin foam model
\cite{SF} (in the models \cite{generalised,iotimgft} the labelling
of the spin foam 2-complex is slightly more involved). The inverse
is also true: any local spin foam model can be obtained from a GFT
perturbative expansion \cite{mikecarlo,laurentgft}. The sum over
Feynman graphs gives then a sum over spin foams, and equivalently
a sum over triangulations, augmented by a sum over algebraic data
(group elements or representations) with a geometric
interpretation, assigned to each triangulation. This perturbative
expansion of the partition function also allows for a perturbative
evaluation of expectation values of GFT observables, as in
ordinary QFT. In particular, the transition amplitude (probability
amplitude for a certain scattering process) between certain
boundary data represented by two spin networks, of arbitrary
combinatorial complexity, can be expressed as the expectation
value of the field operators having the same combinatorial
structure of the two spin networks \cite{laurentgft, iogft}.

$$
\langle \Psi_1\mid\Psi_2\rangle = \int
\mathcal{D}\phi\,O_{\Psi_1}\,O_{\Psi_2}\,e^{iS(\phi)} =
\sum_{\Gamma/\partial\Gamma=\gamma_{\Psi_1}\cup\gamma_{\Psi_2}}\,\frac{\lambda^N}{sym[\Gamma]}\,Z(\Gamma)
$$
where the sum involves only 2-complexes (spin foams) with boundary
given by the two spin networks chosen.

\ \

The above perturbative expansion involves thus two types of sums:
one is the sum over geometric data (group elements or
representations of $G$) entering the definition of the Feynman
amplitudes as the GFT analogue of the integral over momenta or
positions of usual QFT; the other is the overall sum over Feynman
diagrams. We stress again that, in absence of additional
restrictions being imposed on the GFT, the last sum includes a sum
over all triangulations for a given topology and a sum over all
topologies.

\subsection{A peculiar quantum field theory (still, a proper field theory!)}
In the end, {\bf GFTs are a peculiar type of quantum field
theories}, defined on specifically chosen group manifolds. The
main reasons why they are rather peculiar, from a purely
field-theoretic perspective, are:

\begin{itemize}

\item the way in which field arguments are paired in the
interaction term, which makes them a sort of {\it combinatorially
non-local field theories};

\item the resulting combinatorial structure of Feynman diagrams,
given, as we discussed by fat graphs dual to simplicial complexes,
but also presenting no true vertex of interaction, in the usual
QFT sense of simultaneous identification of more than two
configuration variables, and constituted only by \lq loops \rq
(closed lines of propagation of the individual field arguments)
and \lq bubbles\rq (3-cells bounded by several such loops);

\item the fact that all the arguments of the field are naturally
treated on equal footing; if a specific time parameter can be
identified among the group coordinates, still there would be one
such parameter for each argument of the field, thus D in total,
leading to a sort of \lq multi-time dynamics\rq; in the
Hamiltonian analysis of GFTs \cite{iojimmy}, this implies the need
for a polysymplectic canonical formulation and has several
interesting consequences;

\item the fact that, for GFTs characterized by kinetic functions
formed by differential operators, there is then naturally one such
operator for each argument of the field, and a product structure
of the full kinetic term, reproducing again this independent
propagation of field arguments, but also producing technical
complications.

\end{itemize}

However, as for the rest, we have an almost ordinary field theory,
in that we can rely on a fixed background metric structure, given
by the invariant Killing-Cartan metric on the group manifold (or
extensions of it), a fixed topology, given again by the topology
of the group manifold, the usual splitting between kinetic
(quadratic) and interaction (higher order) term in the action, and
the usual conjugate pictures of configuration and momentum space.
This allows us to use all usual QFT techniques and language in the
analysis of GFTs, and thus of quantum gravity, even though we
remain in a background independent (in the physical sense of \lq
spacetime independent\rq) context. The importance of this, in a
non-perturbative quantum gravity framework, should not be
underestimated, we think, and it is at the roots of the strategy
we will propose later on to tackle the issue of the continuum and
semi-classical approximation.

\section{Group field theory as a common framework for discrete quantum gravity}
{\bf GFTs can potentially represent a common framework for
different current approaches to quantum gravity}, in particular
canonical loop quantum gravity\cite{LQG} and simplicial quantum
gravity formalisms, namely quantum Regge calculus \cite{QRC} and
(causal) dynamical triangulations \cite{CDT}, because the same
mathematical structures that characterize these approaches also
enter necessarily and in very similar fashion in the GFT
framework. {\bf We believe in the need to learn from all of them
in order to solve the remaining challenges towards a complete
theory of quantum gravity, and the GFT formalism may be the most
suitable framework in which the many lessons we can draw from all
of them can be brought together and to fruition.}

\subsection{Convergence of formalisms, structures and languages}
Historically, GFTs can be understood as being born as a
generalisation of matrix models \cite{mm} for 2-dimensional
quantum gravity. This generalisation is obtained in two steps: 1)
by passing to generic tensors, instead of matrices, as fundamental
variables, thus obtaining a generating functional for the sum over
D-dimensional simplicial complexes that was the essence of the
dynamical triangulations approach to quantum
gravity\cite{gross,ambjorn}; 2) adding group structure defining
geometric degrees of freedom. The last step is what turns a tensor
model into a proper field theory. In fact, the first example of a
GFT was the group-theoretic generalisation of 3d tensor models
proposed by Boulatov \cite{boulatov}, corresponding to the $D=3$
and $G=SU(2)$ case of (\ref{BF}). Already at this initial stage,
group field theories allowed a direct contact between simplicial
quantum gravity and what we now call spin foam models \cite{SF},
as the Boulatov model produces Feynman amplitudes given by the
so-called Ponzano-Regge spin foam model. As we have discussed
above, we now know that this is just one example of a very general
result \cite{mikecarlo}: the equivalence between (local) spin foam
models and GFT Feynman amplitudes. In turn, spin foam models
\cite{SF} have been a very active area of quantum gravity research
in the past ten years, for two main reasons. First, one obtains a
spin foam model when considering a path integral quantization of
discrete gravity formulated as a gauge theory. Second, spin foams
arise naturally when considering the dynamics of the kinematical
quantum states of geometry as identified by canonical loop quantum
gravity \cite{LQG}. Indeed, from the LQG perspective, spin foams
represents the histories of spin networks and are thus the crucial
ingredient of any path integral or covariant formulation of the
quantum gravity dynamics in LQG. From both the simplicial and
canonical perspective, a sum over spin foams/triangulations,
weighted by appropriate amplitudes, is a crucial ingredient in
defining the dynamics of the gravitational field: in simplicial
quantum gravity because such sum can compensate the truncation of
geometric degrees of freedom that the restriction to a given
lattice imposes; in LQG, because a complete path integral
formulation of the dynamics needs, in general, a sum over all the
histories between given spin network states. At present, group
field theories are the only known tool to define uniquely and
completely such sum over spin foams. Now let us give a closer look
at how the various ingredients of these various approaches, that
all have historically contributed, with hindsight, to the
development of the group field theory formalism, can be identified
and re-interpreted within the formalism itself.

\subsection{Loop quantum gravity and group field theory}
We have mentioned already the first and most basic link between
the group field theory formalism and loop quantum gravity: {\bf
boundary states of generic GFTs are spin networks}, i.e. what has
been identified by the canonical loop quantization programme as
the kinematical quantum states of geometry \cite{LQG}. {\bf The
GFT field itself}, as we have seen, {\bf is interpreted as the
result of a 2nd quantization of a spin network wave function.}
This correspondence can be made more precise, and one can in fact
show \cite{ioeteraSN} that a generic spin network wave function
can be re-expressed as a direct analogue of a multi-particle wave
function, with the particle degrees of freedom being associated to
the spin network vertices; a standard second quantization
procedure applied to these multi-particle wave functions, then,
leads to a field defined on the same group manifold from which
spin network data are taken, and that can be straightforwardly
identified with the GFT field. {\bf GFTs therefore define possible
dynamics for these quantum states of geometry, in a 2nd quantized
formulation,} and in a way that identifies the basic dynamical
degrees of freedom as those associated to the vertices of the spin
networks themselves, that in turn have been shown in LQG to
correspond to elementary chunks of space volume. From these
kinematical considerations, it immediately follows that any
quantum operator that can be defined in the 1st quantized LQG
setting has a 2nd quantized GFT counterpart, that can be, at least
in principle, identified. More importantly, this suggests that the
LQG {\it dynamics} can be embedded and studied within the GFT
setting. There are two equivalent ways in which this can be done.
First of all, as in any QFT, the GFT classical action should
encode the full 1st quantized dynamics, and the classical
equations of motion should correspond to the full dynamical
equations of the 1st quantized wave function. Solving the GFT
classical equations, then, means identifying non-trivial quantum
gravity wave functions satisfying {\bf all} the quantum gravity
constraints, an important and still unachieved goal of canonical
loop quantum gravity, except in some simplified situations. The
same classical equations of motion can be solved, implicitly, also
at the level of the perturbative Feynman expansion: one could
consider the restriction of the GFT perturbative expansion given
above to {\it tree level}, for given boundary spin network
observables \cite{laurentgft}. This is the GFT definition of the
canonical inner product between two spin network states. The
definition is well posed, because at tree level every single
amplitude $Z(\Gamma)$ is finite whatever the model considered due
to the absence of infinite summation (unless it presents
divergences at specific values of the configuration/momentum
variables). Moreover, it possesses all the properties one expects
from a canonical inner product \cite{laurentgft}. This means that
the physical Hilbert space for canonical spin network states can
be constructed starting from the above definition of the inner
product. This shows a concrete example of how the dynamics of spin
network states can be encoded covariantly in a sum over spin
foams, in the same sense in which the dynamics of canonical
gravity in ADM variables can be formulated, in principle, as a
covariant path integral over geometries (see \cite{alex} for more
details on this perspective on spin foam models).

There are of course many open issues regarding the exact
connection between the LQG and the GFT frameworks. One concerns,
for example, the role of spatial topology change. Its status
within LQG is not obvious at present: on the one hand, LQG being
the result of a canonical quantization on a globally hyperbolic
manifold, one would expect spatial topology change to be ruled out
almost by definition; on the other hand, the resulting quantum
states of geometry unavoidably describe also quantum spatial
geometries with degeneracy points, and thus seem to admit the
possibility of branchings of space at those points). In GFT, as we
have seen, non-trivial topologies appear in perturbative expansion
as soon as one goes beyond tree level, and there is no known
mechanism to either suppress or avoid them. Another open issue is
the interpretation, from the quantum gravity point of view in
general, and within LQG in particular, of the GFT coupling
constant; for some proposals on this, we refer to the literature
\cite{iogft}. One more unsettled point is whether one should
expect a direct link between the GFT and the LQG dynamics, i.e.
between the GFT action and the LQG Hamiltonian constraint already
at the level of the, supposedly, microscopic definition of the GFT
itself, or at the level of some macroscopic, effective QFT action
defined starting from the microscopic GFT dynamics. After all, the
Hamiltonian constraint operator of LQG is obtained by a direct
quantization of continuum (and possibly effective) General
Relativistic dynamics, and while one can be lucky enough to
capture some kinematical properties of the microscopic description
of a system, in general one should not expect to capture the exact
microscopic {\it dynamics} of the same starting from some
effective macroscopic description \cite{volovik}, although it is
certainly a possibility. More specific open issues concern the
exact choice of the gauge group, which is usually the full Lorentz
group in GFTs and the $SU(2)$ subgroup in LQG, the need for the
GFT restriction on the valence of spin network vertices, etc
However, while it is clear that much more work is needed to
explore and settle these issues, their presence does not spoil or
modify drastically, we think, the above general picture of the
GFT-LQG relation, and most importantly, all these issues can be
tackled {\it within the GFT formalism itself}.

\subsection{Simplicial quantum gravity and group field theory}
The GFT Feynman diagrams, as we have seen, identify simplicial
complexes to which the GFT assigns geometric data, weighted by
quantum amplitudes that can be related to path integrals for
simplicial gravity on the given complex, and indeed share the same
interpretation. These Feynman diagrams/simplicial complexes are
summed over to define the GFT partition function in perturbative
expansion, and thus the full dynamics.

$$ Z\,=\,\int \mathcal{D}\phi\,e^{iS[\phi]}\,=\,\sum_{\Gamma}\,\frac{\lambda^{N_\Gamma}}{sym[\Gamma]}\sum_{\{J_i\}}\,A_\Gamma(J_i). $$

The relation between GFTs and traditional approaches to discrete
quantum gravity is therefore clear, at least in its general
features. {\bf For given Feynman diagram, and thus fixing a single
triangulation as a discrete model of spacetime, the GFT provides a
quantization of gravity in the spirit and language of quantum
Regge calculus}, by an assignment of geometric data that are (more
or less direct) analogues of the edge lengths used there, and
summing over all such possible assignments. The full amplitude
weighting such assignments, i.e. the specific function of the
geometric data to be used, is specified uniquely b the specific
GFT model one is considering. Schematically:
$$ Z\,=\,\int \mathcal{D}\phi\,e^{iS[\phi]} \curvearrowright \,
 Z_{QRC}\,=\,\sum_{\{J_i\}}\,A_\Gamma(J_i) \approx \lq\lq \int
 \mathcal{D}g \,e^{iS_{GR}(g)} " $$

{\bf If, instead of fixing the triangulation, i.e. considering a
specific GFT Feynman diagram, one freezes GFT field degrees of
freedom (thus fixing the geometric data) to some constant value,
the same GFT provides a definition of the dynamics of quantum
geometry} via a sum over triangulations weighted by purely
combinatorial amplitudes, i.e. functions of the combinatorics of
the simplicial complexes only. This is a definition of quantum
gravity {\bf in the same spirit and language of the dynamical
triangulations approach.} Schematically:

$$ Z\,=\,\int
\mathcal{D}\phi\,e^{iS[\phi]}\curvearrowright
Z_{DT}\,=\,\sum_{\Gamma}\,\frac{1}{sym[\Gamma]}\,A_\Gamma
(\lambda)\approx \lq\lq \int Dg\, e^{i S_{GR}(g)} " $$

The quantum amplitudes weighting histories of the gravitational
field are given, in both quantum Regge calculus and dynamical
triangulations, by the exponential of the Regge action for
discrete gravity, while in most spin foam models the connection
between the quantum amplitudes and the Regge action is clear only
in a particular regime and even there it is rather involved
\cite{SF}. However, such relation is much clearer in the recent
GFTs of \cite{iotimgft}, whose amplitudes have indeed the form of
simplicial gravity path integrals, with clearly identified
classical simplicial gravity actions. GFTs can then be said to
incorporate both traditional simplicial quantum gravity
approaches, and to do so in a nice complementary way. We do not
know, however, is they also do it correctly or whether, by doing
so, they extend the definition of both beyond what is useful or
needed. Much more work is needed, for example,  to study in
greater detail the (classical and quantum) simplicial geometry
corresponding, for given triangulation, to the known GFTs. And
much more work is needed in order to understand what is the QFT
meaning of many of the configurations, e.g. those corresponding to
non-trivial spacetime topologies, or the non-manifold-like ones,
appearing in the perturbative GFT sum over triangulations; how one
could gain control over them is an open, important issue. Also, in
the modern {\it causal} dynamical triangulations approach, the
nice result (that we are going to discuss in the following)
concerning the continuum limit of the sum over triangulations seem
to depend on specific {\it causality} restrictions on the class of
triangulations summed over \cite{CDT}; whether and how one can
understand and implement such restrictions from a field theory
perspective and within the GFT setting is presently unclear. At
the same time, there is hope that the sum over triangulations may
provide a more powerful alternative to the refinement procedure of
Regge calculus to lift the restriction to a fixed simplicial
complex, and that the additional field-theoretic data and
associated gauge symmetries and non-perturbative information of
GFTs can be useful not only because they provide the theory with a
well-identified space of states etc, but also for gaining control
of the sum over triangulations of the dynamical triangulations
approach \cite{GFTopenmath}. To summarize, even given the present
limited level of understanding of GFTs, it is clear that they
represent a unification and a generalisation, that can perhaps
turn out to be useful in the future, of both quantum Regge
calculus and dynamical triangulations, together with a radical
change of perspective on them: GFTs define the 2nd quantized
description of the dynamics of fundamental simplicial building
blocks of space, and simplicial quantum gravity path integrals
arise in a perturbative definition of this dynamics around the
vacuum, either when considering single virtual interaction
processes, i.e. single Feynman diagrams (quantum Regge calculus),
or the full perturbative Feynman sum restricted to its purely
combinatorial properties (dynamical triangulations).

\subsection{Cui prodest?}
So far so good. All this may be interesting and indeed it is
intriguing to speculate of a unifying framework for all discrete
quantum gravity approaches, that encompasses loop quantum gravity
structures as well as simplicial quantum gravity ones. But is it
useful? Can it be helpful in solving any of the outstanding open
problems that these various approaches face? Does it really offer
a new perspective on them and on quantum gravity in general?

In fact, we believe it does offer such a new perspective and that
because of this it can be very useful in helping to solve some of
the current open problems, also by providing new technical tools
for doing so. We have mentioned already some of the possibilities,
e.g. the issue of the dynamics and of the definition of the
physical inner product in LQG, or a possible grasp on non-trivial
topologies in dynamical triangulations. However, what we have
mainly in mind is the issue of the continuum limit, because it is
here that the change in perspective offered by GFTs can be most
relevant. We are going to discuss this issue at length in the
following. Here, we limit ourselves to sketch very briefly what
this change in perspective amounts to and what new tools it
suggests and provides.

The change in perspective, with respect to all the other
approaches we have mentioned, stems from the following
consideration: {\bf all of them, spin foam models, quantum Regge
calculus, dynamical triangulations, arise in perturbative
expansion around the \lq no-particle fundamental vacuum\rq, as
Feynman amplitudes or Feynman diagrams sums.} This means two main
things: 1) that, {\bf from the point of view of GFTs, the
discretization of spacetime used by all of these approaches in
describing the dynamics of geometry, and encoded in a 2-complex
(spin foams) or in a simplicial complex (simplicial quantum
gravity) is not a regularization of the theory (gravity, here) in
the usual lattice gauge theory sense, but corresponds to
describing the physics of \lq few-particles\rq} (be them spin
network vertices or simplices) and virtual processes, with no
individual meaning themselves, except in very limited and specific
approximations; 2) that, at the same time, {\bf the GFT formalism
is in principle suited for going beyond this regime and describe
the many-particle as well as the non-perturbative physics} of the
same system, that is, unless -all- of these approaches are wrong,
quantum spacetime.

Together with a change in perspective, luckily, comes therefore
the possibility of using new mathematical tools and physical
ideas, provided as well by the GFT formalism. This, as said, is a
2nd quantization of the same basic kinematical (space) structures
used in the other approaches, and we know very well how
advantageous it is to have at one's disposal a 2nd quantized and
field-theoretic framework for studying the dynamics of a physical
system described in terms of \lq particle-like objects\rq. A 2nd
quantized, field theory description allows: to overcome the
supreme impracticability of solving the 1st quantized equations of
motions involving many particles (here, very complex spin networks
or extended triangulations), to deal in an easier way with the
symmetries and statistics of the fundamental quanta, to have full
control on quantum (e.g. self-energy) effects. Most importantly, a
field theory description is the best way of: studying the
properties of systems with many degrees of freedom (and, again,
gravity in general, and complex spin networks or extended
triangulations, are certainly examples of such systems);
connecting microscopic many-particle physics and macroscopic,
collective dynamics of the same, its statistical mechanics and the
corresponding thermodynamical quantities. We are going to expand
on this point in the following.

\section{Building up a coherent picture of quantum spacetime}
Once we have seen how (the basic ingredients of) different
discrete approaches to quantum gravity are incorporated within the
group field theory formalism, we can take a fresh look at the many
important results obtained in them, regarding the classical and
quantum nature of gravity and spacetime, and try to re-interpret
them in the GFT language and framework. We are going to be rather
brief, and possibly superficial, in our attempt to summarize in a
few key points what we have learned during many years of quantum
gravity research in such diverse directions, due to space (and
time) constraints, as well as our limited knowledge. We apologize
for this.  This exercise has two purposes. 1) It may help in
acquiring a new understanding of the insights the different
approaches provide, and in analyzing their mutual compatibility,
and possibly also suggests ways in which what we have learned from
one approach can contribute to solving presently open problems of
another or common to all. 2) It is needed in order to check
whether a single coherent picture of quantum gravity, patching
together all these various insights and results, is possible,
within the GFT setting. If it turns out that, indeed, it is
possible, then we believe it would be arguably the best thing to
use it and develop it further.

\subsection{Insights from loop quantum gravity and spin foam models}
So, what have we learned about quantum gravity from loop quantum
gravity \cite{LQG} and spin foam models \cite{SF}? We have learned
first of all that the kinematical degrees of freedom of quantum
space can be captured and encoded in discrete, purely
combinatorial and algebraic structures, spin network states:
graphs labelled by group representations. And this applies as well
to kinematical semi-classical states approximating continuum
geometries. Of this space of states we have strong mathematical
results concerning inner products, kinematical observables,
functional properties and much more \cite{LQG}. Moreover, although
all this has been discovered by a direct canonical quantization of
continuum classical General Relativity with Einstein-Hilbert
action, we now understand this result as a very generic feature of
any description of geometry based on: 1) diffeomorphism invariance
and background independence, requiring a purely relational
description of space, hence the purely combinatorial substratum;
2) a formulation of geometry in terms of connections (and local
reference frames), i.e. a gauge-theory-like formulation of
gravity, hence the use of group elements and representations to
encode gravitational degrees of freedom. These are purely
kinematical considerations, referring solely to the way
information about space and its geometry can be encoded, to a
\lq\lq possible backbone\rq\rq of any theory of quantum gravity,
and thus may well hold regardless of specific dynamical details,
e.g. choices of action, additional symmetry requirements,
spacetime dimension, etc. Similar considerations apply to the
dynamics of space, that can as well be represented in purely
combinatorial and algebraic terms. We have learned this already
from the quantization of the Hamiltonian constraint in LQG
\cite{LQG}, but this is all the more evident in the spin foam
description \cite{SF} of the dynamics of quantum space. As we have
seen, we have again purely combinatorial structures (2-complexes)
labelled by purely algebraic data (group representations and
elements) to represent possible histories of geometry, at the
quantum level. And again, this general features follow naturally
from the requirements of background independence and from a
description of gravity as a gauge theory, either imported from the
canonical formulation, or implemented in some discrete
re-formulation of lagrangian quantum gravity, or somewhat implicit
in categorical quantizations of geometry, which are the main ways
in which a spin foam formalism arises \cite{SF}. Recent results
have confirmed that a spin foam formulation indeed is capable of
describing key properties of the dynamics of quantum gravity, both
in 3 and 4 dimensions, including matter coupling and graviton
propagation, at least in the approximation in which the relevant
spacetime geometry information can be encoded in discrete
structures. For these results, we refer to the literature (see,
e.g. \cite{SF,laurentlibro,graviton}). And also the canonical LQG
formulation of the dynamics has been shown to provide very
interesting physical insights on quantum geometry, at least in the
symmetry reduced context of Loop Quantum Cosmology \cite{LQC}.

From the overview of the GFT formalism that we have given earlier,
it should be clear that all these insights are are not only
compatible but also already fully incorporated in the GFT
framework. In this context, they imply the following: 1) that {\bf
GFT quantum multi-particle states encode correctly quantum
geometric degrees of freedom in a very precise sense, at least at
a kinematical level, and satisfy the requirements of background
independence}; 2) that {\bf GFTs are also able to describe the
corresponding multi-particle dynamics, at least in the
approximation in which the whole perturbative series needs not be
re-summed or high order Feynman diagrams can be neglected.} In
particular, the results on the coupling of matter Feynman diagrams
to spin foams \cite{laurentlibro} show how natural it is to treat
matter Feynman diagrams on the same footing as spin foams, i.e.
GFT Feynman diagrams, which is also confirmed by the corresponding
GFT formulation of the same gravity+matter models
\cite{iojimmymatter}. And the nice results on graviton propagator
in LQG/spin foams \cite{graviton}, using as well and in a crucial
way GFT techniques, seem to us to indicate that GFTs (as LQG and
spin foam models) permit first of all to re-formulate perturbative
gravity questions in a fully background independent language
(which it is we believe the greatest achievement, so far, of this
line of work), and also that GFT perturbative particle dynamics
can in fact reproduce general relativistic semi-classical dynamics
in the (semi-classical, large distance and close to flat)
approximation in which discrete gravity is directly applicable:
GFT few particle physics, and where, in particular, GFT Feynman
amplitudes reduce to semi-classical quantum  Regge calculus, which
indeed is at the heart of these results \cite{graviton}, together
with LQG semi-classical kinematical states.

\subsection{Insights from quantum Regge calculus}
Let us then turn then our attention to what we have instead
learned up to now from (quantum) Regge calculus, referring to the
literature for more details \cite{QRC,hamber}. The main lesson, we
believe, is at the classical level: Regge calculus represents a
beautiful and faithful discretization of classical geometry and of
its dynamics. It has been shown, in fact, that classical Regge
calculus reproduces General Relativity in the continuum
approximation in at least two main ways: 1) the Regge action
approximates well the Einstein-Hilbert action (and the
correspondence generalises to higher-derivatives extensions of the
same) in the sense of measures, and 2) solutions of the linearised
Regge equations converge to analytic solutions to the linearised
Einstein's equations, when some appropriate conditions are met.
Even more confidence in the correctness of the Regge
discretization of classical geometry stems from the possibility of
identifying characteristic symmetries of continuum gravity in the
simplicial setting, including diffeomorphisms, when appropriately
defined, as well as the related discrete Bianchi identities (but
see, on this, \cite{renatediscrete}. In the GFT language, this can
be re-phrased by saying that, {\bf for GFT models that possess
Feynman amplitudes of the form of simplicial path integrals for
(some version of) the Regge action, or in the approximation in
which such form is obtained, there is evidence that the \lq\lq
classical dynamics\rq\rq of the GFT particles can correctly
reproduce relevant features of classical gravity, including
symmetries, and better and better the more GFT particles we
consider.} This already hints at the relation between continuum
geometry and the thermodynamic limit in GFTs (large number of
particles), on which we will say more in the following.

At the quantum level, the results are also interesting \cite{QRC}.
In particular, in the semi-classical, large scale, and flat
approximation, quantum Regge calculus reproduces very well the
graviton propagator and thus Newton's law, plus quantum
corrections, even for simple triangulations. It is quite natural
to expect this to be the case also in GFTs with a simplicial path
integral form of the Feynman amplitudes, and indeed the mentioned
results on the spin foam propagator of the lattice graviton seem
to confirm it, while at the same time confirming the correctness
of the choice of boundary states operated in that context. Many
other results concern matter coupling, quantum cosmology, etc. As
for the definition of the full gravitational path integral in
quantum Regge calculus, the situation is more controversial, and
much debate in particular has focused on the issue of the quantum
measure to be used \cite{QRC}. More precisely, the object of
interest is the continuum limit of the discrete path integral
defined by Regge calculus, on which there are interesting but not
fully conclusive results \cite{hamber}, and about which we will
say more in the next section. As explained above, this discrete
path integral is nothing more (for specific GFT models, or in
special limits of the same) than the GFT Feynman amplitude for a
particular interaction process of GFT quanta.

\subsection{Insights from matrix models and dynamical triangulations}
In matrix models for 2d quantum gravity and in their
higher-dimensional extensions, i.e. tensor models, as well as in
the strictly related dynamical triangulations (DT) approach
\cite{mm,CDT}, the goal is to obtain a consistent and computable
definition of the gravitational path integral, i.e. of the sum
over geometries for given spacetime topology, with some results
being obtained also on the limited extensions of the same to
non-trivial topologies. As such, the classical simplicial geometry
is of limited interest, and indeed it cannot be fully captured by
the approach due to the truncation of the geometric degrees of
freedom associated to the individual lattices. The classical {\it
continuum} geometry, on the other hand, is possibly reproduced to
the extent in which the DT partition function reproduces the
gravitational continuum path integral. {\bf In GFT terms} this is
easily understood, as it this means that, {\bf once one has frozen
the field degrees of freedom, the classical particle dynamics}
(classical simplicial gravity) {\bf cannot be reproduced in a
satisfactory manner, but at the same time the continuum field
dynamics} (continuum quantum gravity) {\bf could still in
principle be reproduced, at least to the extent in which the
truncated sum over Feynman diagrams, restricted to its
combinatorial properties, reproduces properties of the full field
partition function.} Therefore, all the many results obtained in
this approach refer to the continuum approximation of the discrete
gravitational path integral defined as a sum over triangulations,
and we defer their discussion to the next section. Here we limit
ourselves to notice that work in matrix models and dynamical
triangulations has resulted in an immense amount of results and
available tools, both analytical and numerical, in an almost
complete understanding of 2d quantum gravity with a nice
discrete-continuum correspondence, in both Riemannian and
Lorentzian cases, and in important results obtained recently for
higher dimensions concerning this discrete-continuum
correspondence, in the Lorentzian context of so-called {\it
causal} dynamical triangulations \cite{CDT}.

\section{The problem of the continuum: current strategies from a GFT perspective}
{\bf Given our favorite formulation of quantum gravity, using
discrete structures of some sort to describe spacetime and to
encode quantum geometric degrees of freedom, does it reproduce, in
some controlled and well defined approximation, a smooth
spacetime, and is the quantum dynamics of spacetime geometry
effectively approximated, in the same regime, by continuum General
Relativity, possibly modified by quantum effects?} This is the
problem of the continuum in quantum gravity, for how we see it.
And this is, in our opinion, {\it the} outstanding unsolved issue
that all the current approaches to quantum gravity, and certainly
the ones we have mentioned, loop quantum gravity and spin foam
models, quantum Regge calculus, dynamical triangulations, have to
tackle hard and solve, to be considered successful. The same, of
course, is true for group field theory. The importance of
obtaining a satisfactory understanding of this issue cannot be
overstated, we believe, as it would amount to showing that our
favorite formalism, whatever it is, does indeed provide at least
{\it one possible} quantum theory of gravity. In absence of such
result the connection with gravity would remain a (more or less
plausible) hypothesis, and, as stressed, for example, in
\cite{renate}, any interpretation of the discrete expressions one
has in terms of quantum spacetime structures can be taken only as
a suggestion, {\it before} a physically correct continuum
approximation to them has been found. {\bf The group field theory
formalism, in the perspective we are proposing, can offer new
tools to solve this issue} to each of the different approaches it
(potentially) subsumes, and at the same time capitalise on their
results and insights. However, we believe that {\bf it also calls
for a change in perspective and for a consequent new strategy.} We
will be arguing in this direction in the next section; here we
would like first to briefly overview the strategies currently
adopted within the other approaches, all of course sensible and
potentially successful, and then \lq\lq translate\rq\rq them in
the GFT language, since this translation will make clear why a
change in perspective and strategy is naturally suggested.

\subsection{The loop quantum gravity/spin foam strategy}
Research on the semi-classical and continuum approximation in loop
quantum gravity and spin foam models has been mainly carried out,
at least in the 4-dimensional setting, in the canonical
formulation and is mainly confined to the kinematical
setting\footnote{The exceptions, that may come to mind, are the
many results in Loop Quantum Cosmology \cite{LQC}, and the recent
progress on the spin foam calculation of the lattice graviton
propagator. However, the first apply to symmetry reduced
situations, where it is possible to encode {\it all} the (finite
number) degrees of freedom of the continuum theory in the discrete
spin network structures. The second is limited to perturbative
physics {\it around a semi-classical space geometry}, first of
all, but, more important, remains confined at the level of
(justified) discrete approximations and large scale information,
thus not really addressing the issue of the continuum in this
framework.} The starting point of the LQG/SF strategy (using
$SU(2)$ spin networks and related observables) for recovering
continuum physics is the construction of appropriate kinematical
quantum states of space which approximate continuum space
geometries in some sense. The first type of such
semi-classical/almost continuum states are the so-called \lq\lq
weaves\rq\rq \cite{weave,weavestat}. These are defined by a
(directed) graph embedded in a reference compact space $\Sigma$,
the links are dressed with holonomies of an $SU(2)$ connection in
the representation $j=1/2$, with appropriate intertwiners
labelling the vertices of the graph. This graph is taken to be a
huge collection of loops in $\Sigma$, uniformly distributed with
respect to some classical 3-geometry $h_{ab}$. The mean spacing
between the loops (akin to a sort of lattice size for the graph)
is of the order of the Planck length $l_P$. This means that the
number of loops is approximately $N=(\frac{L}{l_P})^3$ where $L$
is the distance scale corresponding to the volume region one is
interested in approximating, measured in the reference metric
$h_{ab}$. The observables considered as a probe of the
semi-classicality of our quantum state are areas of surfaces in
$\Sigma$ and 3-volumes of regions contained in it. The nice result
is that for large enough volume regions, the areas and volumes as
computed quantum mechanically on the weave state are very close to
the ones measured in the classical continuum metric $h_{ab}$, and
with very small uncertainties. In this sense, one can say that the
quantum state considered has a good continuum and semi-classical
approximation \cite{weave}. This type of construction can be
extended to consider random weaves and averages over ensembles of
graphs, using statistical techniques \cite{weavestat}. A different
type of improvement of this construction is to change test
observables \cite{coherent}, using for this scope the basic
canonical pair of variables of loop quantum gravity, i.e. triad
and holonomy operators. The resulting quantum states are then
semi-classical {\it coherent states} providing expectation values
for both of them that are close to the classical continuum values,
as well as minimizing the uncertainties of both, in an appropriate
sense. The resulting quantum states are then an even more
satisfactory approximation of continuum 3-geometry, and many nice
results can be proven for them \cite{coherent} (overcompleteness,
Ehrenfest properties, etc). Notice however that we are still
confined at the kinematical level, while what we are really
interested in reproducing, starting from out quantum gravity
formalism, is the continuum {\it dynamics} and the {\it spacetime
continuum}. The way to do this test in the Hamiltonian/canonical
setting would be to study the action  of the Hamiltonian
constraint of the theory on these weave or coherent states. This
is extremely complicated, due also to the intrinsic complications
involved in the very definition of the Hamiltonian constraint
operator, and has not been done, to the best of our knowledge.
More work has been devoted recently to the spin foam formulation
of the dynamics, so maybe one would want to use these weave or
coherent states in that context. It has not been done, yet. The
general idea however would be to use the above
semi-classical/almost continuum states as boundary states for an
appropriate spin foam model and compute the quantum gravity
analogue of 2-point functions between two of them; the spin foam
amplitudes would impose the quantum dynamics and the result should
then be compared with continuum path integral
calculations\footnote{One would also have to compute observables
other than 2-point functions, but this does not alter our
argument.}. The calculation could be done for fixed spin foam
2-complex, but most likely should involve a sum over spin foams,
that could then be truncated because of some physical
requirements. One way to define such sum would be through the
corresponding GFT formulation, with the GFT here used only as a
auxiliary tool, devoid of physical meaning, for generating the sum
over 2-complexes. All this is possible and sensible. However,
notice the orders of magnitude that would be involved, generally
speaking, in such calculations: if we aim at reproducing continuum
physics over a scale of, say, $L=10^{-19}cm$ (the distance scale
of a quark), we would need boundary states, in our spin foam
calculations, that are weaves with about $N=10^{42}$ loops, or,
which is arguably the same, spin networks with a similar number of
vertices. The complexity of the spin foam complex would go
accordingly. It is not obvious that such a calculation would be
doable, and at the very least we are lead to look for some
alternative, more efficient procedure.

Let us look at the GFT translation of the same procedure, taking
the GFT formalism to be physically meaningful in itself and not
just a mathematical tool, and see how the above sounds like. In
the GFT language, interpreted {\it realistically}, the procedure
would then be the following: 1) consider a carefully chosen
multi-particle state of a given GFT (a quantum field theory)
corresponding to a wave function satisfying some carefully
specified conditions (with respect to your favorite choice of test
observables); this state should contain about $N=10^{42}$ GFT
quanta, say; more precisely you should consider two of these
states, one per boundary in a typical \lq\lq scattering\rq\rq
process; 2) construct the corresponding field observable and
insert it in the GFT partition function; 3) expand the GFT
partition function in perturbative expansion around the vacuum
state (i.e. the state with no GFT quanta), i.e. in Feynman
diagrams; these Feynman diagrams give all the possible virtual
interaction processes of the $10^{42}$ initial and final
particles, including all quantum loops, self-interactions etc.;
even for the simplest diagrams (e.g. tree level and next to tree
level), their complexity will be of the same order of and scale
with the complexity of the boundary states; 4) compute the
transition amplitude in this Feynman expansion, maybe truncating
the expansion to some given order in the GFT coupling constant
(notice that the needed order would be necessarily extremely
high).

{\bf The strategy is not wrong, in any sense, but it definitely
does not look like what one would naturally do to study the
physics of such hugely populated multi-particle state in a field
theory context.} The basic point is that, {\bf when we choose as
our system of interest a hugely populated particle state, we put
ourselves immediately in the situation in which the vacuum
no-particle state and its physics is not relevant, the Feynman
diagrams of the individual particles are not relevant, in a sense
the microscopic dynamics itself is not relevant
anymore}\footnote{One can of course be more optimistic and hope
that a smooth continuum spacetime arise, and a general
relativistic description of it, holds already, say, for distances
100 times the Planck length; this will make the number of needed
particles $N=10^6$. The numbers are then vastly different but the
result is the same: for this number of quanta, the direct solution
of the corresponding microscopic dynamical equation for wave
functions or the study of their dynamics via Feynman expansion
around the vacuum are at best unpractical and possibly even
conceptually mistaken. If such a lucky situation occurs, it would
simply mean that already at the order of $10^6$ particles, we are
free to take the limit $N\rightarrow\infty$.}. {\bf In any case,
the Feynman diagrammatics and the individual particle picture is
not the most convenient language to describe the relevant physics
of these states.} We are lead to look for an alternative.

\subsection{The quantum Regge calculus strategy}
In quantum Regge calculus\cite{QRC,hamber}, the theory is defined
by the Euclideanized (or statistical) discrete gravity path
integral on a fixed lattice (most often hypercubic, then
subdivided into simplices) $T$ (thus also for fixed topology,
usually the sphere or the torus): \be Z_T=\prod_e\int
\mathcal{D}l_e\, e^{- S_{Regge}(l_e)}, \ee where $e$ labels the
edges of the lattice, $l_e$ are the corresponding edge lenghts,
which are the fundamental variables, integrated over with some
measure $\mathcal{D}l_e$, and the most studied version of the
discrete action, in 4d, is the Regge one augmented by quadratic
higher derivative terms (a discretization of the Riemann tensor
squared):

\be S_{Regge}(l_e) = \sum_{t} \left( \lambda V_t(l_e) - k A_t(l_e)
\epsilon_t(l_e)+ a \frac{(A_t(l_e)
\epsilon_t(l_e))^2}{V_t(l_e)}\right), \ee where the sum runs over
the triangles of the 4d simplicial complex, $A_t$ are their areas,
$\epsilon_t$ the associated deficit angle (discrete curvature),
and $V_t$ is the contribution of the given triangle to the total
4-volume of the lattice \cite{QRC,hamber, renatediscrete}. The
partition function is then a function of the coupling constants
$\lambda$ (cosmological constant), $k$ (inverse of Newton
constant), and $a$. The integration over the edge lengths is
usually cut-off both in the IR and UV, to ensure convergence.

Studying the continuum approximation of this theory means studying
the above partition function and appropriate geometric observables
(average curvature, average square curvature, etc) for very large
simplicial lattices (often at fixed total 4-volume) in a scaling
limit, while removing the cut-offs, as a function of the coupling
constants. The aim is to show that in a region of the parameter
space the above reproduces continuum spacetimes and continuum
geometric observables, thus representing a good (regularised and
computable) substitute of the formal continuum gravity path
integral. As said, this analysis has been done exclusively for
statistical path integrals over euclidean geometries, and mainly
numerically. The main results are the evidence for a two-phase
structure: for a certain $k_c$ the average curvature vanishes; for
$k> k_c$ (small $G_N$) the simplicial complex degenerates into a
crumpled phase incompatible with a smooth geometry with simplices
of very small volumes and large curvature; for $k<k_c$ there is
instead evidence for a smooth phase, depending also on the value
of $a$ and $\lambda$, with small (and negative) curvature. See
\cite{hamber} for more details. There is evidence for a second
order nature of the phase transition \cite{hamber}, which is what
one needs in order to have long-range correlations, but this
evidence does not seem to be considered fully conclusive by the
community (see, e.g. \cite{renatediscrete}). The result seems to
be rather generic, i.e. not too strongly dependent on the specific
measure $\mathcal{D}l$ chosen or on the specific topology or
lattice structure chosen, even if for irregular lattices the phase
structure is more involved (more critical points) and singular
structures seem to appear (spikes) and the choice of measure
becomes more important. In the end, we cannot yet conclude whether
this approach reproduces continuum physics or not, but we
definitely have gained lots of insights in the properties of
similar discrete gravity path integrals, and many tools to analyze
them have been developed.

Once more, this does not look like the most natural procedure to
adopt to study the continuum approximation of the same structures
when embedded and re-interpreted in a GFT context. The discrete
gravity path integral on a fixed lattice, in fact, amounts to the
evaluation of a single GFT Feynman amplitude for a given
interaction process of the GFT quanta, and all the lattice
prescriptions used in quantum Regge calculus require a Feynman
diagram with about $10^3-10^4$ vertices of interaction (numerical
simulations have been performed with up to $16^4\sim 6 \times
10^5$ lattice size, and the continuum approximation is expected to
be only improved going to larger lattices). {\bf Such a huge
Feynman diagram computation would indeed capture some information
of the many-particle physics of the corresponding GFT, which is
again suggested to be the regime corresponding to continuum
gravity, but the truncation to a single Feynman diagram is most
likely not consistent within the GFT setting. Moreover, just as in
LQG, it seems that to study the many-particle dynamics of the
theory at the level of perturbative expansion around the vacuum is
 definitely not the most convenient thing to do.}

\subsection{The dynamical triangulations strategy}
In the traditional euclidean triangulations programme, the theory
is defined by the partition function:

\be Z=\sum_{T}\frac{1}{C_T}\,e^{- S_{Regge}(l, k, \lambda)}, \ee
i.e. by a sum over {\it equilateral triangulations} $T$, at fixed
topology (usually the spherical one), with fixed edge length $l$
(which is interpreted as a cut-off), weighted by a symmetry factor
(the automorphism group of the triangulation, $C_T$, and a
euclideanized exponential of the same Regge action usually limited
to a cosmological and a curvature term, thus in the end a function
of the combinatorics of the triangulation only and of the
parameters $l$,$k$,$\lambda$. The continuum approximation involves
again evaluating explicitly this sum $Z(\lambda,l,k)$ or, more
precisely, its Legendre transform $Z(N_4,l,k)$ which corresponds
to work for fixed number of 4-simplices $N_4$ and thus with fixed
4-volume $V \sim l^4 N_4$. Having done this, one is interested in
the thermodynamic limit $N_4\rightarrow \infty$, $l\rightarrow 0$,
$V \sim$ constant. Simplifying a bit, the resulting phase
structure was found to be given again by two phases separated by a
critical value of $k$, $k_c$, depending ont he volume $N_4$. For
$k<k_c$ we have a crumpled phase characterised by small curvature,
high graph connectivity and very large Hausdorff dimension. For
$k>k_c$ one finds an elongated phase with large and positive
curvature and an effective branched-polymer geometry with
effective Hausdorff dimension equal to 2. One could still hope
that a continuum theory is defined at the transition point, if the
transition was second order, but further analysis (again not fully
conclusive) suggested that the transition is instead first order.
For more details and further references, see
\cite{renatediscrete}.

The situation changes drastically in the modern form of this
approach, the so-called {\it causal dynamical triangulations}, to
the point that one can even make the case
\cite{CDT,renate,renatediscrete} for the {\it origin} of the
troubles encountered in the euclidean dynamical triangulations, as
well as, to some extent, in euclidean quantum Regge calculus, in
finding a good continuum approximation to be the dominance of
pathological configurations such as baby universes and other types
of singular geometries. These configurations are basically
unavoidable in the euclidean setting. They are not so, however, in
a Lorentzian one, where instead one can indeed identify conditions
on the triangulations summed over that rule out them from the
start (i.e. by construction). This is what is achieved in the
causal dynamical triangulations approach. Here the basic
ingredients for the construction and definition of the
triangulations summed over are (see \cite{CDT} for more details):
1) a local light cone structure, i.e. a differentiation between
spacelike and timelike edges (which have a relative
proportionality factor $\Delta$ for their values, on top of the
difference in the sign of their square); 2) the existence of a
global discrete time function; 3) no spatial topology change
allowed with respect to this \lq time\rq structure. The
triangulations are then weighted by a complex exponential of the
same Regge action but now for Lorentzian simplicial geometries.
The results are striking \cite{CDT}. There are now three phases:
a) for large $k$ a phase characterised by 3-dimensional slices of
a branched-polymer type, so not a 4d smooth geometry, once more;
b) for small $k$ and small asymmetry parameter $\Delta$ a phase
with crumpled 3-dimensional slices, similar to the euclidean
setting; so, again, not a smooth 4d geometry; c) for sufficiently
small $k$ and sufficiently large $\Delta$, a stable, extended
4-geometry, with Hausdorff dimension equal approximately to 4 and
a global shape of spacetime related to a simple minisuperspace
model of gravity, similar to those used in quantum cosmology. This
is strong and exciting evidence for a smooth geometry and thus a
continuum limit, even though several features of the model itself
and of the resulting dynamics of geometry are yet to be
understood, such as whether the results are robust with respect to
limited extensions of the ensemble of triangulations considered,
how much exactly of the full dynamics of general relativity is
recovered in the continuum approximation, whether there is a way
to generate analytically the above sum over triangulations, that
is at present constructed algorithmically and only studied
numerically, etc.

How does the GFT translation of the above sound like? In the GFT
language, the above corresponds to the following: 1) consider a
specific GFT model, producing Feynman amplitudes with appropriate
exponential form (either real or complex) for a discrete gravity
action (with field theoretic data interpreted as either euclidean
or lorentizian discrete geometries); 2) fix all the field
theoretic data, e.g. the momenta of the GFT field to some constant
value, giving then equilateral triangulations dual to the GFT
Feynman diagrams (producing the parameters $l$ and, in the causal
case, $\Delta$); this corresponds to restricting to a specific
momentum regime for the GFT particles, i.e. to particles all
having the same momentum; 3) restrict the perturbative sum over
Feynman diagrams to only those diagrams of some given topology
(and further restrictions have to be in place to recover the
causal restrictions of \cite{CDT}); finally, perform the {\it
whole} sum computing in this way the corresponding restricted
sector of the theory partition function, and appropriate
observables. Once more, we see that {\bf one necessarily needs to
study Feynman diagrams or arbitrary combinatorial complexity, and
involving huge numbers of GFT quanta, supporting further the idea
that continuum physics corresponds to the many-particle physics of
the theory.} Importantly, the work on dynamical triangulations
provide lots of technical tools for studying it. {\bf The CDT
results,} moreover, {\bf seem to indicate that, at least in that
regime of the GFT, some continuum physics can indeed be captured
in satisfactory form by this procedure}, which is exciting indeed.
However, once more the above procedure seems not so convincing
from the GFT perspective (that of course one is free not to take):
first of all it is well possible that the DT and the CDT
restrictions at the level of GFTs are not consistent from a field
theory perspective. Here we are not so much concerned by the
restriction on the momenta, which may well simply correspond to a
particular sector of the GFT, and thus to a reasonable
approximation of the full theory. Rather, what may be more
problematic is the restriction to fixed topology and, in the CDT
case, to fixed slicing structures of the diagrams summed over. We
have a too poor understanding of the GFTs themselves
\cite{GFTopenmath} to specify what these restrictions amount to,
from a purely field theory perspective. For example, we may run
into problems in asuming these restrictions, if they do not
clearly amount to a classical limit, say, as for example the
large-N planar limit of matrix models, and still involve removing
the GFT analogue of quantum loops or the like. Modulo these
remarks, {\bf it is clear that the (C)DT restriction does indeed
amount to extract at least some non-perturbative information far
beyond the physics of the GFT vacuum state, i.e. the few particle
physics, so it is a sensible thing to do even within the GFT
setting; in practice, in fact, amounts to solving the theory
(computing the partition function) at least in a restricted
sector, which may well turn out to be the one in which continuum
gravitational physics lies.}

{\bf However}, the same doubt put forward concerning the other
approaches applies: {\bf how convenient is it to study the
non-perturbative many-particle physics, and the corresponding
vacuum state and its dynamics, using what remains a perturbative
expansion around the no-particle vacuum, encoding the the
many-particle dynamics in hugely complicated Feynman diagrams, and
then re-summing all of them?} Again, the GFT perspective calls for
the use of different tools and for a change in strategy.

\subsection{Lessons and further motivations}

Let us summarize briefly the outcome of this sketchy overview.
First of all, {\bf {\it all} these strategies and approaches {\it
do} teach us something about GFTs}, when embedded in it. Second,
{\bf {\it all} of them suggest or strongly support the view that
continuum physics corresponds to the many-particle sector of the
GFT formalism}, and most likely involve collective and
non-perturbative effects. Third, {\bf their translation in the GFT
language suggests that maybe, although we have been indeed
studying the relevant sector of the (GFT) theory, we have not used
the most convenient set of tools} and language for doing so.
Fourth, luckily enough, {\bf the GFT formalism potentially
provides us with all the non-perturbative, field-theoretic tools
and concepts for trying out a different strategy.} As we had
stressed, in fact, we know from condensed matter physics and
statistical physics that field theory and 2nd quantization
language are the most convenient ones to study many-particle
physics, the corresponding phases, collective behaviour, etc.

\section{Quantum spacetime as a condensed matter system}
Let us now give some more specific suggestions for what this GFT
perspective seems to imply. We will put forward an hypothesis for
the continuum phase of a GFT, i.e. the phase or regime of the
theory in which we expect continuum gravitational physics to be
reproduce, and some general hints at what the strategy to check
this hypothesis could be. The general idea for the above will be
to take GFTs seriously for what they (formally) seem to be, at
least as a working hypothesis, and consider them as the
microscopic description of a very peculiar condensed matter
system, which is quantum space. In other words, {\bf we will
consider the GFT quanta}, that can be pictured, as we have seen,
as spin network vertices or (D-1)-simplices, {\bf as the true
\lq\lq atoms of quantum space\rq\rq , its fundamental hypothetical
constituents, and the GFT formalism as the microscopic
(fundamental?) formulation of their quantum dynamics}, thus
described in terms of a peculiar (non-local, etc) quantum field
theory, but a quantum field theory nonetheless. Then, we will
broaden the discourse a little and try to summarise some of the
general insights that, once we have taken this standpoint, come to
us from condensed matter physics (and from condensed matter analog
gravity models).

\subsection{If GFT is its microscopic description, what is the
continuum and how to get there?}

We have seen that all current approaches seem to suggest that
continuum gravitational physics is obtained in what is, in the GFT
language, the (very) many-particles sector of the theory. This
perfectly match the working hypothesis of GFTs as the microscopic
description of the atoms of space. In other words, we most likely
need a {\bf very large number} of them to constitute a region of
space that can be governed effectively by continuum gravitational
physics, and be described by a continuum space to start with.

Moreover, from results in these approaches, as well as from
general physical intuition and, again, from the perspective of
spacetime as a \lq\lq material\rq\rq of some sort, made of (GFT)
constituents, we would expect these many constituents making up
the continuum to be {\bf very small}, in the appropriate sense,
probably of order of the Planck length (volume). In GFT terms,
generally speaking, this translates into the {\it GFT quanta to be
in a low momentum regime}. On top of this, we expect the quantum
constituents of space to be {\bf governed, in their continuum
phase/regime, by collective dynamical laws}, not anymore by the
microscopic individual dynamics, simply because otherwise we would
have noticed already the \lq\lq true\rq\rq atomic nature of space.
Finally, whatever the exact phase looks like, whatever the
symmetries characterizing it are, and whatever the effective
dynamics governing it is, we expect our condensed matter system,
i.e. quantum spacetime, modelled by our favorite GFT, to be {\bf
very close to equilibrium}. In other words, we expect a continuum
description of spacetime to prove itself correct, and not only
possible, when close to an {\it equilibrium} and stable
vacuum/phase of the (GFT) system, at least to scales close to the
sector of the physical world that has already probed (by us).
Again, this is simply because otherwise we would have most likely
already noticed a failure of the continuum description of
spacetime. From this perspective, the breakdown of general
relativistic theories of geometry in cosmological situations or in
black hole physics can be {\it speculated} to be a sign of a phase
transition occurring in the (fundamental?) GFT system.

Notice that none of the above implies that the continuum
approximations goes necessarily in hand with the semi-classical
approximation, which may be needed later on to simplify/extract a
specific dynamics or for capturing some relevant features of the
system in the regime we are interested in, but as far as the above
reasoning is concerned, the possibility of {\bf a continuum
description may even be {\it the result of a purely quantum
property of the system}}. We will give later on an explicit
proposal for this.

So, {\bf a continuum space is a very large number of very small
GFT quanta very close to equilibrium, i.e. very close to some yet
to be determined many-particle vacuum, to be described
collectively and whose dynamics is to be given by continuum
larger-scale equations}. This seems (to us, at least) just a
description of a fluid (whether gaseous or liquid or what else, is
to be determined by hard, technical, future work), close to
equilibrium, governed indeed by hydrodynamical equations.

The picture that seems to come out of the above reasoning, then,
and more indirectly (we admit that) from work in the various
approaches to discrete quantum gravity we have discussed is that
of {\bf {\it quantum spacetime as a (quantum) fluid of GFT
particles, governed microscopically by the GFT partition function,
but macroscopically by a suitably identified GFT effective
hydrodynamics.}}

As we had stressed, this is at present just a suggestion, of
course, given the little we understand GFTs themselves and the
(basically nihil) amount of work that has been devoted up to now
to develop and test it. But we find it a very intriguing and, most
important, convincing one. It immediately implies one thing: at
least for a while, at least from the GFT standpoint, and only if
we intend to tackle the issue of the continuum approximation and
its effective dynamics, it may be convenient to partially forget
about spin foams and even the simplicial gravity description of
the GFT system, and focus our attention on other aspects of the
formalism. This is simply because, as we have stressed, the
perturbative formulation of GFTs, which is where the spin foam and
the simplicial gravity descriptions appear, is very useful for the
physical interpretation of the system, of its quanta and field
theoretic data (indeed, we have relied exclusively on it for all
of the above reasoning), but it is {\it technically} useful for
describing the system in its few-particle regime. If we are
interested in describing the many-particle behaviour of the same
system we should move away from the no-particle vacuum.

In its stead, we need to develop first and then use a {\bf {\it
statistical group field theory}} formalism for identifying first
and then select the different phases of the theory, i.e. the
possible equilibrium configurations in which the system may find
itself, hoping that some of the GFT models we have or we will
construct for the scope allow for the existence of at least one
with the properties that allow for a continuum geometric
description. Second, we need to obtain an {\bf {\it effective
field theory or hydrodynamic description}}, coming from the
fundamental GFT, for describing the dynamics close to the
different phases, and probably tied to each particular phase under
consideration. We will speculate more, but also try to be more
specific, about how both may look like in the next section.

\subsection{What can quantum gravity learn from condensed matter theory and analogue gravity models}
The idea of spacetime as a condensed matter system in general, and
as a  fluid in particular, and of GR as an hydrodynamic effective
description of it, is of course not new and has been advocated
many times, and very convincingly, in the past
\cite{jacobson,hu,volovik,laughlin,padmadaran}, and is both
motivated by and an inspiration for the many condensed matter
analog gravity models \cite{analog}. What is new here is only the
argument that it is the very research in non-perturbative quantum
gravity carried out to date, and the many results obtained in the
many approaches it is split into, that points in this direction.
Also, what is new here is the hypothesis that GFTs can represent:
1) the framework in which these many approaches to quantum gravity
and their insights can be seen as part of a single coherent
formalism and physical picture of spacetime; 2) also because of
this, a solid and motivated formalism to be used to realise
concretely, in mathematical and physical terms, the suggested idea
of spacetime as a condensed matter system of a peculiar type, and
a concrete, if tentative only, description of its microscopic
structure. This description, moreover, as we stressed repeatedly,
uses a field theory language that may facilitate the application
in this context of traditional condensed matter ideas and tools
(probably suitably adapted).

This is probably the main contribution that GFTs can provide
researchers working in condensed matter analog gravity models:
{\it a concrete formalism and system} on which to apply their
insights, if they are interested in unravelling the true
microscopic structure of quantum spacetime, and not only in
finding out more about its effective continuum description, once
interpreted as a condensed matter system, or in using the same
gravitational analogy and the general relativistic tools to
discover more interesting properties of the usual condensed matter
systems (Bose-Einstein condensates, etc)\footnote{Needless to say,
both things are definitely worthwhile and of fundamental
significance; simply, they are not quantum gravity issues.}.

In other words, it is often stated in the analog gravity
literature and in the condensed-matter-but-interested-in-gravity
community that \cite{hu,volovik,analog}: 1) quantum gravity is not
so much about quantizing general relativity in a strict sense, but
rather about identifying the microscopic constituents of space and
provide a tentative description of their microscopic dynamics; 2)
we do not know what this microscopic structure and dynamics is; 3)
the current top-down approaches to quantum gravity are so
different and so complicated that no coherent picture  and no
clear indication about the fundamental structure of space is
provided by them, that could serve directly for the application of
the insights coming from condensed matter theory. What we have
argued is the following. The first thing is true, and {\bf it is
the very same approaches to non-perturbative quantum gravity that
have lead (in a rather tortuous way) to the GFT formalism which
itself is {\it not} a quantization of classical GR} (just look at
the GFT action). The second statement is true in a sense, {\bf we
do not have a clear and complete picture of the spacetime
microscopics}, but false or at least overly pessimistic in
another: {\bf we have several candidates for this microscopic
structure, and one, the GFT formalism, that seem to encompass many
of them}. The third statement is false: {\bf the respective
pictures that at least some of these approaches to quantum
gravity} (those we have discussed) {\bf provide are not only
compatible and coherently build up a tentative picture of quantum
spacetime} (the one encoded in the GFT formalism), {\bf but also
one that allows for a rather direct application of condensed
matter concepts, formalisms and techniques for understanding the
microscopic-macroscopic and discrete-continuum transition.}

What quantum gravity, and in particular the GFT approach, can
learn from condensed matter (CDM) physics and from condensed
matter analog gravity models is much more.

Concretely, the main help that condensed matter techniques can
provide stems from the fact that {\bf in that context}, as
stressed in \cite{analog}, {\bf the transition from discrete
microscopic physics and continuum macroscopic one is well
understood} conceptually and there are many theoretical tools that
can be applied to its analysis and study. As we have seen, this is
the main open problem of the discrete quantum gravity approaches
have to solve, even after they have provided a tentative
description of microscopic spacetime. This holds for GFTs as well,
and its field theory setting makes the application of CDM
techniques even more straightforward. More generally, taking a CDM
perspective means also a conceptual shift with respect to what we
expect from our theory, and how we approach our physical
challenges. We list here only some of the CDM wisdom (for more,
see \cite{hu,volovik,laughlin,anderson,analog}), that is useful
for approaching our quantum gravity problems, in our opinion.  We
should not expect a rigorous, deductive path from the microscopic
dynamics to the macroscopic one, and even the kinematics (relevant
variables, symmetries, etc) at the macroscopic scale or in the \lq
continuum phase\rq , thus in the hydrodynamic regime, can be very
loosely related to the one of the corresponding microscopic
theory. In other words, even the relation between microscopic
variables and collective ones is often less that direct, and the
specific form of the microscopic QFT for your atoms is often not
at all similar to the macroscopic effective QFT for the resulting
fluid. In particular, many of the small details of the microscopic
theory become irrelevant at the hydrodynamic effective level. This
is governed mainly by general macroscopic symmetries and
associated conservation laws, that should acquire thus a
fundamental importance in our model building. It is not
reasonable, in light of the above, and at least if one is first of
all interested in showing that a continuum approximation exists,
to demand necessarily for exact treatments or to look for exact
solutions of microscopic dynamics, because this exactness will
almost inevitably end up being irrelevant at a different (larger)
scale. All this should apply to our future treatment of the GFT
formalism, in our attempt to use it to obtain the correct
macroscopic effective continuum description. Nothing revolutionary
here, of course, but things that are worth keeping in mind in
quantum gravity research, and in particular when one sees
spacetime as a condensed matter system, because they are often
neglected (by us, at least). Also, we are warned that experimental
is absolutely crucial for guiding model building and for guessing
what are the relevant features in the hydrodynamic regime, thus at
the effective level. The recent development of quantum gravity
phenomenology \cite{giovanni} it therefore of extreme importance,
also in this condensed matter interpretation.

At the same time, as stressed very nicely in \cite{anderson} (see
also \cite{laughlin}): \lq\lq The behavior of large and complex
aggregates of elementary particles, it turns out, is not to be
understood in terms of a simple extrapolation of the properties of
a few particles. Instead, at each level of complexity entirely new
properties appear\rq\rq . This is a warning but also an
encouragement because it implies richness and potential fun in
unravelling it.

\section{Guessing the future: several research directions, an hypothesis and some speculations}
The above discussion has been very general, serving only the
purpose of sketching what are further inputs to the GFT
perspective on the continuum we are advocating, again, as a
working hypothesis. Now we will try to be a bit more specific
about how one can develop further and what may come out of this
condensed matter perspective, in concrete terms, in the GFT
framework. We will put forward one specific proposal for what can
be the phase of the GFT, i.e. the relevant vacuum for the GFT
multi-particle physics, where continuum geometry and its dynamics
could be reproduced, and then explore, tentatively, some
possibilities for the dynamics of the theory in this phase, and
how it can relate to known formulations of classical continuum
gravity.

It should be clear that, given our present understanding of the
GFT formalism, any guess in this direction can be only partially
based on known results, but rather speculative. The study of the
GFTs in their own right, treated as peculiar but {\it bona fide}
field theories, is in its infancy and only the first basic steps
have been or are being taken \cite{iojimmy}. Nevertheless, they
already provide some hints of what may come next, and we are going
to build upon these hints in the following.

Before we do so, let us mention three other directions of work
that, in the perspective we are advocating, are certainly relevant
(see also \cite{GFTopenmath} for a more detailed discussion). One
if the development and use of renormalization group techniques.
The renormalization group is in fact one of the most powerful
tools we have in field theory and in condensed matter physics to
explore the structure and behaviour of our system at different
scales. It is indeed applied routinely in condensed matter for
investigating phase structures, which is exactly what we have
argued we have to do in our GFTs. In particular, we believe that
it would be very important, and of great direct relevance for
solving the problem of the continuum, to develop the formalism of
the Wilsonian Exact Renormalization Group for group field
theories, with the construction of the effective action and the
analysis of the corresponding flow, for specific GFT models. This
would not only prove the consistency of the given models
(renormalisability, etc) but also suggest what is the relevant
form of the theory (action) at the scales we expect to be related
to continuum physics. A second one is the study of classical
solutions of the GFT equations. Of course, they encode
non-perturbative information about the system, and thus are also
relevant for the continuum phase. This work has started
\cite{elainstantons}. However, we would like also to stress that,
from a condensed matter point of view, it may be even more
important to construct {\it approximate} solutions to the GFT
dynamics, tailored to the multi-particle situation. The third, and
maybe most important, is the analysis of the GFT classical
symmetries, to be done both at the lagrangian and hamiltonian
level \cite{iojimmy}; this is because, as stressed, macroscopic
behaviour and hydrodynamics in particular are likely determined
more by these symmetries, or their broken version, than by the
exact microscopic GFT dynamics.

\subsection{Geometrogenesis using GFTs}
Our proposed general scheme for the emergence of continuum
geometry from the dynamics of the GFT quanta can be seen as a
particular possible implementation of the {\it geometrogenesis}
idea.

This is the catchy name given in \cite{ltf} to a conjectured phase
transition of a combinatorial and algebraic model of quantum space
described by a a labelled graph, much alike spin networks, between
a high-temperature \lq pre-geometric phase\rq in which space has
the form of a complete graph, and thus no notion of locality or
geometry (e.g. distance), to a \lq geometric phase\rq in which the
graph acquires a more regular, local structure, where geometric
data can be identified. Furthermore, the data labelling the graph
 then allow for the emergence of matter degrees of freedom, having
the role of qausi-particle moving on the resulting regular
lattice, in the same way as the model of topological order studied
by Wen et al \cite{wen} does, in terms of string condensation.

Now, the details of the model do not concern us here. We just want
to note the similarity with the idea we are proposing for the
emergence of the continuum in GFTs. The basic quantum states of
the GFTs, as we have seen, are characterized by labelled
combinatorial structures as well, of the spin network type (or,
dually, of a simplicial type). It seems to us that because of
this, any phase transition in a GFT setting will be described by a
transition from some irregularly structured and labelled graph or
from an ensemble of such graphs to a more regular and ordered one
at lower temperatures, in the same spirit as the model of
\cite{ltf}. Further, we are suggesting that after the ground state
has been identified its own effective dynamics will be described,
if the scenario we are suggesting is correct, by an effective
continuum field theory with a geometric interpretation, and in
principle derivable (but not necessarily deducible) from the
microscopic GFT. Both the hamiltonian function driving the
transition, and thus the selection of the ground state lattice,
and the effective hamiltonian governing the dynamics of
quasi-particles around the resulting ground state, the two main
ingredients of the model in \cite{ltf}, can in principle be
derived from any given choice of GFT action, whose dynamical
content is indeed the same, after appropriate simplifications. If
our understanding is correct, then, the model of \cite{ltf} can be
interpreted as an effective simplified GFT Hamiltonian, and
similar models can be constructed and inspired by the GFT
formalism as well. Conversely, we believe that more work in the
direction opened by the model \cite{ltf} will be of importance
also for the research programme we are suggesting, in that it will
amount to explore models that may indeed capture relevant features
of GFT phase transitions and vacua as well.

\subsection{Spacetime as a condensate: a GFT realisation?}

\subsubsection{Continuum space as a GFT Bose-Einstein condensate}
Our tentative proposal for a relevant vacuum of a GFT model in
which a continuum approximation could be expected, i.e. a
continuum and geometric phase of the model, is a simple one: a
Bose-Einstein condensate. Again, here it is not so much important
the idea in itself, because the similarities between continuum
spacetime and condensates have been noticed long ago and a similar
possibility has been advocated by several authors, and very
convincingly \cite{hu,volovik} and the effective (and emergent)
spacetime character of real Bose-Einstein condensates (those
stored in laboratories) is the basis of many condensed matter
analog gravity models \cite{analog}. What is important here is the
fact that the concrete realization of this scenario within a
specific microscopic model of quantum spacetime, i.e. a GFT model,
seems to us not only possible, but within reach. Of course, such
scenario involves first of all the development of a statistical
group field theory formalism, the identification of the GFT
analogues of relevant thermodynamical quantities, and more, and,
as we have noticed above, even basic steps in the analysis of GFTs
apart from their Feynman amplitudes have been taken only recently
\cite{iojimmy}. We will now sketch, also based on these initial
results, how thermodynamical quantities in a GFT setting could be
defined and then how the possibility of a Bose-Einstein condensate
of GFT quanta could be realized, including some likely features of
the resulting vacuum state.

GFT thermodynamic quantities \cite{iostat} will have to be defined
in a formal way, letting ourselves be guided, at first, only by
the field theory look of the GFT formalism, and only in a second
stage one should try to match the definition of each of them with
a corresponding physical interpretation. In turn, this physical
interpretation will have to rely almost exclusively on the
(pre-)geometric interpretation that the GFT variables have in the
context of the Feynman expansion, i.e. in the context of
simplicial gravity. This can be done more easily in an Hamiltonian
setting, and in the same context we will give now a sketch of a
possible concrete definition of Hamiltonian (thus of a GFT \lq\lq
energy\rq\rq) and temperature, while for other quantities we can
only offer guesses, at this point, although reasonable ones, we
hope.

Consider a GFT action like (we restrict here to the free theory,
which sffices for our present purposes)\cite{generalised}:

$$ S=\left( \prod_i \int_G dg_i
\int_\mathbb{R} ds_i\right) \,
\phi^\dagger(g_1,s_1;...;g_D,s_D)\prod_i\,\left( i\partial_{s_{i}}
+ \square_i \right)\phi(g_1,s_1;...;g_D,s_D) + h.c.$$ with $g_i\in
G$, $s_i\in\mathbb{R}$, $\square$ being the Laplace-Beltrami
operator on $G$, for generic group $G$ (Riemannian or Lorentzian).
The kinetic term has the structure of a product of differential
operators, each acting independently on one of the D (sets of)
arguments of the field. Each of them is a Schroedinger-like
operator with \lq \lq Hamiltonian\rq\rq $\square$. This suggests
that one should consider the variables $s_i$ as \lq\lq time\rq\rq
variables, to be used in a GFT generalization of the usual
time+space splitting of the configuration space coordinates, with
the group elements treated instead as \lq\lq space\rq\rq. This
implies that we have a field theory with D \lq\lq times\rq\rq, all
to be treated on equal footing. The approach chosen in
\cite{iojimmy} is to use the DeDonder-Weyl generalized Hamiltonian
mechanics, as developed at both the classical and quantum level as
a {\it polysymplectic (or polymomentum) mechanics} by Kanatchikov
\cite{kanatchikov}, as a starting point and to adapt it to the
peculiar GFT setting.

The general idea is the following \cite{iojimmy}. One starts from
a \lq\lq covariant\rq\rq definition of momenta, hamiltonian
density, Poisson brackets, etc treating all \lq\lq time
variables\rq\rq on equal footing at first, i.e. when defining
densities. Then one defines 'scalar' quantities referring to each
\lq time direction\rq (to be turned into operators at the quantum
level), including a set of D Hamiltonians, by integration over
appropriate hypersurfaces in $(G\times\mathbb{R})^{\times D}$, so
that each Hamiltonian refers to a single time direction, but at
the same time all time directions are treated equally but
independently. A similar procedure is adopted for other canonical
quantities, e.g. Poisson brackets, scalar products etc.

Let us sketch one example of such procedure, for the case $D=2$,
referring to \cite{iojimmy} for more details. We start from the
naive phase space $(\phi,\phi^\dagger,\pi_\phi^i=\frac{\delta
L}{\delta\partial_{s_i}\phi},\pi_{\phi^\dagger}^i=\frac{\delta
L}{\delta\partial_{s_i}\phi^\dagger})$, with the product structure
of the kinetic term resulting in a peculiar expression for the
momenta, e.g. $\pi_\phi^1 = (-i\partial_2
+\square_2)\phi^\dagger$, and define the DeDonder-Weyl Hamiltonian
density (summation over repeated indices understood):
$$ \mathcal{H}_{DW} =  \pi_{\phi}^i
\partial_{s_i}\phi + \pi_{\phi^\dagger}^i
\partial_{s_i}\phi^\dagger - L = 2 \pi_{\phi^\dagger}^1 \pi_\phi^2 + i
\pi_\phi^1 \square_1 \phi + i \pi_\phi^2 \square_2 \phi + h.c. .$$
One then proceeds to re-write it as a sum of two contributions,
each uniquely associated to a single time parameter:
$\mathcal{H}_{DW} = \mathcal{H}_1 + \mathcal{H}_2$ , with
$\mathcal{H}_i = \pi_{\phi^\dagger}^1 \pi_\phi^2 + i \pi_\phi^i
\square_i \phi + h.c$. The Hamiltonians governing the \lq time
evolution\rq\ with respect to the different time directions
identified by each variable $s_i$ are then defined by integration
over independent hypersurfaces, each orthogonal to a different
time direction, e.g. $H_1 = \int ds_2 dg_i \mathcal{H}_1$. Each
$H_i$ results in being independent of time $s_i$.

One can then proceed, after suitable decomposition in modes of
fields and momenta, the definition of (a GFT-adapted version of)
the covariant Poisson brackets, etc, to the canonical quantization
of the theory, with the definition of a Fock structure on the
space of states. We refer once more to \cite{iojimmy} for the
results of this analysis.

From the above results, it is easy to guess how the notion of GFT
temperature may be defined, because it simply involves following
the usual QFT procedure. One could repeat the analysis above but
now requiring periodicity of the fields in the $s_i$ variables,
with period $\beta$, and would then be left with a partition
function in hamiltonian form:

$$
Z= \int \mathcal{D}\phi \mathcal{D}\phi^*\; e^{i\sum_i \int ds_i
H_i(\phi,\phi^*) }
$$

with the integration over $s_i$ restricted to the interval
$(0,\beta)$, and thus obtaining, after Wick rotation in the same
$s_i$ variables:

$$
Z= \int \mathcal{D}\phi \mathcal{D}\phi^*\; e^{-\beta \sum_i
H_i(\phi,\phi^*) } =\int \mathcal{D}\phi \mathcal{D}\phi^*\;
e^{-\beta H_{tot}(\phi,\phi^*) }
$$

with $\beta=\frac{1}{kT}$ defining the GFT temperature. The notion
of temperature, then, may be defined, and indeed the corresponding
quantity will play the role of a temperature at least at the
formal level. However, its physical interpretation will have to be
studied with care (even its dimensions may not be those of a
temperature). In other words, just as the variables $s_i$ played
the role of time in the formalism, and could be treated formally
as such in a consistent way, but still do not have the geometric
interpretation of time variables on any physical spacetime, not
even at the simplicial level, similarly the GFT temperature $T$
may be found to correspond, say, at the simplicial level, to a
geometric quantity that a priori has no similar interpretation,
even though the GFT sees it indeed as a temperature parameter. An
even clearer example is the notion of energy in the above simple
GFT. The hamiltonian in each \lq time direction\rq is given by
$\square_i$ acting on the group manifold $G$ for the i-th field
argument, and corresponding to a particular set of field modes
solutions of the GFT equations of motion. In momentum space, i.e.
in representation space, it is given simply by the Casimir of the
group $G$, and for compact groups (Riemannian models) it will have
a discrete spectrum with minimal eigenvalue $0$. Thus we see that
the group representations $J$ correspond to the \lq\lq
energy\rq\rq of the GFT. However, their geometric interpretation
(at least at the simplicial level) is that of (D-2)-volumes, i.e.
distances, areas etc according to the dimension chosen. This is
the type of procedure we were envisaging above for defining
thermodynamical GFT quantities: be guided first by the field
theory formalism, then look for a geometric interpretation. As a
further example, as the GFTs are field theories on the group
manifold $G^{\times D}$, its is (the normalisation chosen for)
this group manifold and any eventual cut-off in the group
integrals that will provide a definition of GFT \lq\lq
volume\rq\rq in which the GFT quanta could be confined. From this
quantities, and the partition function itself, one can proceed to
define other thermodynamical quantities, standard statistical
ensembles etc.

What is most relevant for us here is that within the same type of
formulation, a straightforward proof of Bose-Einstein condensation
seems possible, at least for the free theory, and in the case in
which indeed the GFT quanta are bosons (which is not obvious
\cite{iojimmy}). Indeed, one expect to be able to even adapt to
the peculiar GFT setting the standard (textbook) derivation of the
Bose distribution and proceed as usual. In the model sketched
above, in fact, one expects that for fixed number of particles
(GFT quanta) and at low temperature $T$, the system will reach its
ground state represented by (almost) all the GFT quanta condensed
into the same state $J=0$. Again, according to the simplicial
geometry emerging from GFTs in perturbative expansion, this means
having all (D-2)-volumes being of Planck size. Work on this is
currently in progress \cite{iobec}.

The interpretation of this {\bf vacuum state} is exciting, we
think. It corresponds to {\bf a free gas of spin network vertices
or of (D-1)-simplices that has condensed in momentum space, i.e. a
Bose-Einstein condensate of spin network vertices/simplices;
geometrically, a Bose-Einstein condensate of the fundamental
building blocks of quantum space all of Planck size}.

This also resembles, in general terms, the heuristic picture of a
\lq\lq semi-classical state\rq\rq in LQG, with two differences: no
embedding is needed for its definition, and it is selected \lq\lq
dynamically\rq\rq , in a GFT statistical setting.

From a more general perspective, there are many reasons why a
condensed phase of this kind would be a very attractive
possibility, in our opinion, for the vacuum relevant for the
continuum limit. We have mentioned the first: it is realisable in
concrete terms, and not just an hypothesis. Still at the practical
level: the theory of Bose-Einstein condensates is vast and lots is
known about them (see for example \cite{bose}), so in principle
many tools from the condensed matter theory of BEC systems can be
imported in the GFT setting to study the property of this new
phase. At the theoretical and conceptual level it is also very
attractive: it is {\bf a purely quantum phenomenon}, thus a
realisation of the possibility we anticipated that the emergence
of a continuum spacetime from GFT structures could be considered
indeed a quantum effect; it is {\bf rather generic} \cite{bose},
being robust to the presence of interactions, even strong ones, if
they are repulsive, but surviving (when dealt with much care) also
small attractive ones; it gives rise to a pletora of emergent
phenomena \cite{hu,volovik,analog}; as we will discuss in slightly
more details in the following, {\bf the approximate collective
motion of the condensate admits (in mean field theory
approximation) a description in term of a classical (better, 1st
quantized) equation}, the Gross-Pitaevskii equation; {\bf
condensate atoms move as a whole, so that small purely quantum
effects can be amplified}, and one can speculate the same to
happen for this quantum gravity condensate, thus leading (we are
speculating!) to observables quantum gravity effects or, more
likely, to the possibility that large scale properties of
spacetime (e.g. features of GR) that we are accustomed to, can be
understood as originating from purely quantum features of this GFT
vacuum.

To summarise, we are proposing the possibility that {\bf GFT will
produce geometrogenesis in the form of a condensation of the GFT
particles in momentum space accompanied by the approach to
equilibrium of the system} (otherwise, no hydrodynamic description
is possible).

Let us close this section with a comment, that will be relevant
for the following guesses at the effective dynamics of the
condensate. GFT quanta (think of them now as open spin network
vertices) are labelled by both representations of $G$ and by
corresponding vector indices in the representation spaces. It may
happen (and indeed is what we would expect because of symmetry
considerations at the level of the GFT action) that the GFT
hamiltonian, and thus the energy of the vacuum state does not
depend on these additional parameters. Now, suppose that the
condensation is not complete, so that the vacuum state is actually
a mixture of spin net vertices with $J=0$ and $J=1/2$, for
$G=SU(2)$, or in general of lowest eigenvalue (which has also a
single value for the vector indices) and next to lowest eigenvalue
for the energy. Alternatively, suppose that the lowest eigenvalue
is forbidden by some symmetry or by the quantum measure; or, more
generally, the lowest allowed eigenvalue (for some group $G$ and
choice of GFT action) may have a representation space of dimension
bigger than 1. What this means is that we do not necessarily
expect the condensation to lead to a unique vacuum state, even in
the $T\rightarrow 0$ limit. Instead, it may lead us to any of the
quantum states corresponding to $N$ spin network vertices for the
lowest allowed representation parameters and some given choice for
their vector indices. Now in particular, one can consider all
linear combinations of such states, obtained by contracting in all
possible ways the spin network vertices along their open links
labelled by the vector indices. Each of these possible
contractions, which is equivalent to a gluing of the dual
(D-1)-simplices, corresponds to a possible choice of the topology
of the corresponding quantum space, formed by the same spin
networks/simplices. Of course each possible choice also
corresponds to a different effective condensate wave function
\cite{bose}, that then carries a dependence on the resulting
topology of quantum space. If on the other hand, the GFT dynamics
or some additional symmetry consideration will select a specific
contraction of the vector indices or the absence of any such
contraction, once more this will amount to selecting one specific
space topology for our quantum space in this phase.

\subsubsection{Effective dynamics of spacetime from GFT}

Let us move to discuss how we could try to extract and study the
effective dynamics, actually the hydrodynamics, of the GFT
condensate. In discussing this issue, once more the present status
of the field will force us to remain at the level of arguments,
guesses, speculations. Again, we hope the reader will find them
interesting.

Generally speaking, the effective collective dynamics will depend
heavily on the phase the system is in, i.e. on the vacuum selected
by the GFT microscopic dynamics. At this stage, even to guess it
is impossible. However, we can try to forecast some general
features and ask ourselves very general questions about it.

We are assuming here that a sort of Bose-Einstein condensate has
formed, that the system is at equilibrium or very close to it,
that we have made one specific choice of vacuum state, obtaining a
specific effective vacuum wave function \cite{bose}, or
equivalently a classical field (the order parameter).

It is possible that a clever redefinition of the field variables
will bring us collective variables with a direct geometric
interpretation, say  connection field or a metric, so that we
could hope that the effective hydrodynamics for these collective
variables is given directly by some extended gravity theory.
However, we find this possibility very unlikely, for the following
reasons:

\begin{itemize}
\item while the effective topology of the physical quantum space
is probably determined by the vacuum (following the comments at
the end of the previous section), nothing seems to select for us
the effective topology of {\it spacetime}; in general, we should
expect an effective theory in which spatial topology change and
non-trivial spacetime topologies are included;

\item in analog gravity models \cite{analog}, the effective
spacetime that quasi-particles see may be very different from the
original spacetime on which the microscopic field theory is
defined, in both geometry and topology, but the spacetime on which
the {\it hydrodynamics} is defined is very close to the one one
started from;

\item in particular, the GFT we have started from has the
interpretation of a discrete 3rd quantized formulation of gravity
and indeed, at least in perturbative expansion, produces discrete
virtual spacetimes of arbitrary topology, and moreover it was a
theory on an internal group manifold and not a physical continuum
spacetime; we expect neither the \lq\lq formal level of
quantization\rq\rq nor the nature of the manifold on which the
effective field is defined to change with respect to the original
microscopic (group) field theory.
\end{itemize}

For all the above reasons, and some others, we expect the
effective GFT dynamics for the chosen condensate vacuum to be not
directly of the form of an extended gravitational theory on a
fixed spacetime, but rather of the form of a continuum 3rd
quantized field theory of gravity, i.e. of a quantum field theory
on a continuum superspace (space of continuum geometries). This
type of gravitational theories have not been much studied, beyond
the original definition \cite{giddingsstrominger,mcguigan}, but
are supposed to have the general action (schematically):

\be S = \int_\mathcal{S} \mathcal{D}X \Psi^*(X)
\mathcal{H}(X)\Psi(X)\, + \Lambda \int \Psi^{n}(X) V(X) \ee

where $\Psi(X)$ is a scalar field on the superspace $\mathcal{S}$,
i.e. the space of all space geometries (not spacetime) for given
space topology $\Sigma$, and $X$ are then coordinates on this
space, i.e. some geometric variables (3-metrics, connections,
etc); the (non-local) interaction term $V(X)$ generates, in
perturbative expansion spatial topology changing processes
(producing disconnected universes) while the free kinetic term is
given by a canonical Hamiltonian constraint $\mathcal{H}$. Notice
that the superspace $\mathcal{S}$ is a metric space itself
\cite{dewitt}.

As we have said, for our GFT condensate, we expect the effective
field, call it $\Psi$ as well, to be determined by the vacuum
state, from which would most likely inherit also the choice of
space topology $\Sigma$ and the topological and metric properties
of the effective superspace $\mathcal{S}$, that will depend on the
space topology chosen. In turn, as we have said, the properties of
the vacuum state depend on the original choice of GFT field and of
group manifold $G^{\times D}$. We then expect the emergent
superspace to be some sort of group manifold, with an exact
structure determined by the topology of space we have selected
with the vacuum, and thus again parametrised by group elements or,
equivalently by a (gravity) connection.

To summarise, we would probably obtain, as our effective GFT
hydrodynamics of quantum space, 1st order versions of the old
quantum field theories on superspace. Nothing is known (to the
best of our knowledge) about how these may look like, and a
detailed analysis of such possible field theories (involving the
metric structure of a 1s order superspace, first of all) is called
for.

In general, then, our effective GFT hydrodynamics, in the GFT
analogue of the mean field approximation, will be a continuum
field theory of the form:

\be S = \int_\mathcal{S} \mathcal{D}X \Psi^*(X)
\mathcal{K}(X)\Psi(X)\, + \int V(\Psi,\Psi^*) \ee

for some kinetic term $\mathcal{K}$ and higher order (non-local)
interaction $V(\Psi,\Psi^*)$.

The corresponding equations of motion with be our hydrodynamics
equations, non-linear equations for the field/wave function $\Psi$
that will represent the GFT analogue of the Gross-Pitaevskii
equation for Bose-Einstein condensates \cite{bose}. Notice that
the above field theory can be easily recast in a more customary
hydrodynamic form by redefining the basic variables to $\Psi(X) =
\sqrt{\rho(X)} e^{i \theta(X)}$ where $\rho(X)$ is the condensate
density and $v(X) = \nabla\theta(X)$ is the condensate velocity
field.

Let us now see how the link with continuum GR (in some extended
form, probably) can be investigated. The type of gravity theory we
would have obtained will be encoded, and hopefully fully
specified, by the quadratic term in the above action, that would
give the effective Hamiltonian constraint of the corresponding
canonical theory. Notice that all of the above (and of the
following) is at the level of {\it classical} effective theories.
We then would have to extract the quadratic part of the action,
here represented by $\mathcal{K}$. However, it is clear that the
split of the above action, and more generally the very form of the
effective hydrodynamics action depends strongly on the specific
mean field ansatz one has chosen to obtain it\footnote{As they
say, mean field theory, and in general the procedure of
constructing effective dynamics for collective variables, is a
complicated art.}. Anyway, assuming that, in some approximation,
we have got up to here, we could then compare the kinetic term
$\mathcal{K}$, which would be in general a differential operator
on an effective 1st order superspace $\mathcal{S}$, and thus
depending on connection variables and their conjugate variables,
with the classical Hamiltonian constraints of various canonical
1st order formulations of gravity for space topology $\Sigma$, or
re-interpret it as such, and study in this way what type of
effective gravity theory our GFT reproduces in this phase, i.e.
for this choice of condensate vacuum state\footnote{In principle
it would be also possible to extract the corresponding lagrangian
form for the same gravity theory and even the corresponding
continuum path integral, i.e. the 2-point function for the
corresponding free field theory on superspace. Obviously this
would have only a formal meaning, and limited applicability, just
as the formal quantization of hydrodynamics has, and in any case
will not resembles at all the original GFT we started from, just
as the quantization of hydrodynamics for ordinary quantum fluids
does not reproduce at all the underlying microscopic atomic theory
\cite{volovik}.}.

Another possibility, that we mention en passant, comes from the
interpretation of classical gravity as a single particle theory on
superspace \cite{greensite}. In our case, the continuum superspace
is effective and corresponds to the effective manifold on which
our GFT condensate lives. The procedure for identifying classical
gravity in our hydrodynamic field theory on superspace is
consistent with this interpretation. But what if classical gravity
is a {\it \lq\lq quasi-particle\rq\rq} of the above theory on
superspace, and not a particle? Then the effective superspace it
would live in would not be given by $\mathcal{S}$, but by a space
with an effective geometry function of $\rho(X)$ and $v(X)$
\cite{analog}. We are not going to expand on this, but it is clear
that in this case the body of knowledge developed in condensed
matter analog gravity models \cite{analog} would become even more
directly relevant.

It is clear that the realm of possibilities for the structure of
the vacuum and even more for the way to extract effective dynamics
for it, and to find our what back to classical gravity, is
enormous. This is true even if one accepts the idea of the correct
vacuum being represented by a condensate of the type we suggested.
And there are for sure many other plausible hypothesis that can be
made at this stage. Again, condensed matter physics wisdom
suggests to be cautious because condensed matter systems are rich,
and always richer than we imagine. {\bf We simply wanted to
suggest
 {\it one possible path from the microscopic discrete to the
macroscopic continuum}}: microscopic GFT $\rightarrow$ condensate
$\rightarrow$ condensate hydrodynamics $\rightarrow$ effective
continuum 3rd QGR $\rightarrow$ approximate free theory
$\rightarrow$ classical (extended) GR.

This probably means we have been un-cautious enough already.

\section{Conclusions}
We have presented a brief introduction to the group field theory
formalism for quantum gravity. We have then argued that GFTs may
provide a common framework for several other discrete approaches
to quantum gravity (loop quantum gravity, quantum Regge calculus,
dynamical triangulations), and shown how the connection with these
other approaches can be understood. Having done so, we have tried
to sketch the elements of a single coherent picture of quantum
spacetime, incorporating the insights and results achieved in all
these different approaches, as seen from a GFT standpoint. We have
tried to argue that the GFT formalism offers also a new
perspective on the same structures.

We have then stressed the importance of solving the open problem
of the continuum approximation of the discrete structures
representing spacetime at the quantum level in these quantum
gravity models, including GFTs, and overviewed the strategies
adopted in loop and simplicial approaches to do so, and the
results obtained. At the same time, we have translated these
strategies in the GFT language, showing that the GFT formalism
would suggest a different one instead, and then sketched what we
believe is a new GFT perspective on the continuum problem in
quantum gravity. This amounts to consider quantum spacetime as a
condensed matter system and the GFT as the microscopic quantum
field theory for its fundamental constituents. We have finally
outlined a GFT strategy from tackling the problem of the emergence
of the continuum, put forward an hypothesis for the relevant GFT
phase, a Bose-Einstein condensate, and sketched a (rather
speculative, at present) programme for realizing this idea and
connecting GFT microscopics to continuum gravity and GR, obtained
from the effective hydrodynamics of the GFT condensate.

\medskip

We hope that, in spite of necessary conciseness of the first part
of this contribution, and of the speculative nature of much of the
second, we have managed to elicit interest for the ideas presented
and for this, we believe, very exciting area of fundamental
theoretical physics that is non-perturbative quantum gravity. The
hope is also that the reader will then join the efforts of
researchers working in this area, and contribute to turning the
present speculations into solid results, in the conviction that
most of the many impressive results already obtained in this
fascinating field have been just tentative suggestions or
speculations at an earlier stage.

\section{Acknowledgements}
I would like to thank J. Ambjorn, A. Ashtekar, F. Dowker, L.
Garay, B. Hu, J. Henson, S. Liberati, L-K. Lim, R. Loll, F.
Markopoulou, P. Massignan, P. Machado, C. Rovelli, L. Smolin, M.
Visser, G. Volovik, R. Williams, for insightful and useful
discussions, comments and criticisms. I would also like to thank
the organizers of the Conference \lq\lq From Quantum to Emergent
Gravity: Theory and Experiments\rq\rq, and especially F. Girelli,
for the invitation to participate and for a very interesting and
stimulating conference.

\end{document}